\documentclass[preprint,3p,times,twocolumn]{elsarticle}  

\usepackage{amssymb}
\usepackage{amsmath}
\usepackage{upgreek}
\usepackage{enumerate}
\usepackage{lscape}
\usepackage{rotating}
\pdfoutput=1 
\newcommand*\diff{\mathop{}\!\mathrm{d}}
\usepackage{hyperref}
\usepackage{lscape} 

\usepackage{lineno}

\journal{Nuclear Instruments and Methods A}

\begin{document}

\begin{frontmatter}



\title{The mathematical theory of photomultiplier tube calibration}


\affiliation[a]{organization={Physics Department, National Technical University},
            city={Zografou},
            postcode={15780}, 
            state={Athens},
            country={Greece}}
\affiliation[b]{organization={School of Applied Mathematical and Physics Science, National Technical University},
            city={Zografou},
            postcode={15780}, 
            state={Athens},
            country={Greece}}
            
\author[a]{L. N. Kalousis}\corref{cor1}
\cortext[cor1]{Corresponding author.}
\ead{leonidas.kalousis@gmail.com}
\author[a]{E. A. Dris}
\author[b]{A. Vynias}

\begin{abstract}
In this technical note we describe the main features of the mathematical theory of photomultiplier tube (PMT) calibration.
Attention was paid to explain the various arguments and concepts in a simple and pedagogical manner that everybody understands. 
The basic operational principles of a PMT are discussed from a theoretical standpoint. 
The essential steps of its function (photoconversion, focusing, multiplication, etc.) are laid down together with the mathematical schemes necessary to model the charge output of a PMT when illuminated by a faint poissonian light source. 
In case of important omissions we direct the reader to some of the standard references. 
The most common numerical methods, used to calculate the charge amplification function $S_R(x)$, are also presented.
As an example, we plot $S_R(x)$ utilizing a gamma function model for the single photoelectron (SPE) response and showcase its main characteristics. 
The basic techniques for gain calibration are also introduced, and we probed their precision using toy Monte Carlo data. 
Additionally, data from a Hamamatsu R1408 PMT were analyzed.
Finally, we show how the presence of soft charge component can affect the results of gain determination. 
We conclude this report with some general comments regarding \emph{in situ} calibration of large-scale detectors. 
We hope that this document can serve as a reference for students and young researchers that want to learn more about the theory of PMT calibration.
\end{abstract}




\end{frontmatter}


%

%
%

\section{Introduction}
\label{sec:intro}

\begin{figure}[!t]
\centering
\includegraphics[width=7.0cm, height=4.9cm]{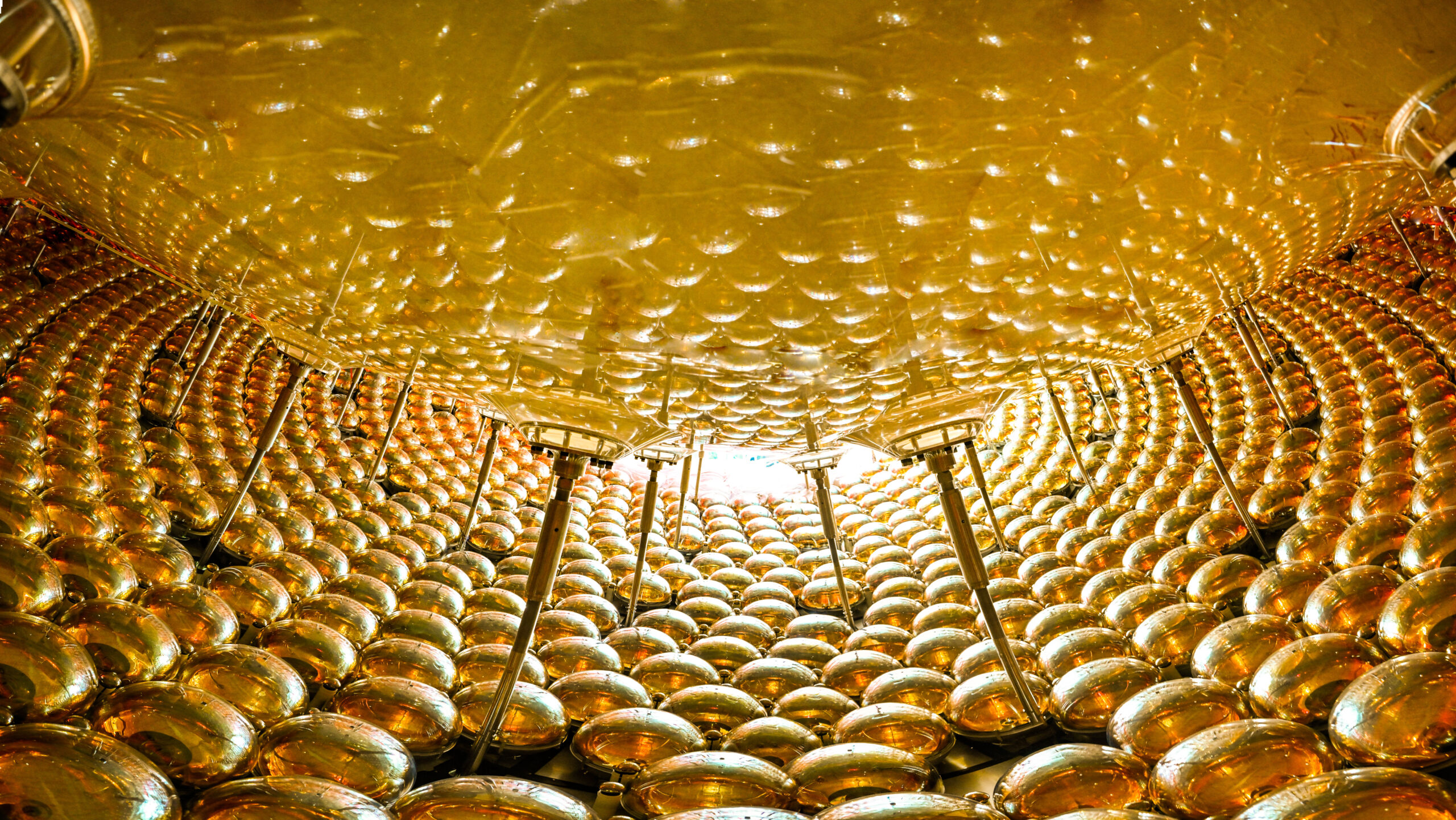} 
\caption{Picture of the JUNO central detector~\cite{infn}. PMTs are shown to be attached at the walls of a stainless steel vessel. }
\label{fig:juno}
\end{figure}

Photomultiplier tubes (PMTs) are devices extensively used in low-energy nuclear physics, particle physics and medical applications~\cite{leo}. 
Their striking stability, simple principles of operation and relatively low cost make them attractive solutions in the instrumentation of large coverage detectors. 
Within neutrino physics, massive monolithic detectors usually employ a sizable number of PMTs to detect optical signals created in the target material. 
For example, the Jiangmen Underground Neutrino Observatory (JUNO) uses approximately 18\,000 twenty inch PMTs to monitor 20 ktons of ultra-pure liquid scintillator, Fig.~\ref{fig:juno}. 
This configuration allows the central detector to achieve an unprecedented level of  roughly $3~\%/ \sqrt{E}$ in energy resolution, necessary for an unambiguous deter\-mination of the neutrino mass ordering \cite{juno}. 
Furthermore, it will be able to pursue a rich physics program ranging from neutrino oscillations to solar and atmospheric neutrinos, geoneutrinos and supernova neutrinos~\cite{juno1,juno2,juno5,juno6,juno7}.
The first result published by JUNO collaboration provides the most precise up-to-date measurement of the solar neutrino oscillation parameters~\cite{res}.

Now, a PMT (as the name suggests) is basically a device that detects light. 
It does so by converting optical signals to an electric current that can be readout by electronics.
It can be made to detect faint light pulses at the level of one photon per pulse. 
Usually the current is integrated in the time domain and a charge value is registered. 
State it differently, for a given amount of light that hits the PMT there is a charge value recorded. 
An important aspect of PMTs operation is the fact that the number of photons that trigger the device is (up to expected statistical fluctuations) proportional to the charge measured at the output. 
Furthermore, in most physics detectors, and especially in scintillation counters, the number of photons produced is proportional to the energy of the incident particle. 
Therefore, a good understanding of PMTs functioning 
is 
rather important for any physics experiment. 

This document seeks to describe the basic working principles of PMTs, together with the mathematical schemes necessary to model their behavior, in a simple and pedagogical language that everyone understands. 
We present the standard machinery for PMT calibration and a comparison study between different techniques is also included. 
Finally, we show how the presence of a soft charge component can affect the results of gain determination.

%

\section{Fundamental concepts}
\label{sec:basic}

\begin{figure}[!t]
\centering
\includegraphics[width=8.5cm, height=4.9cm]{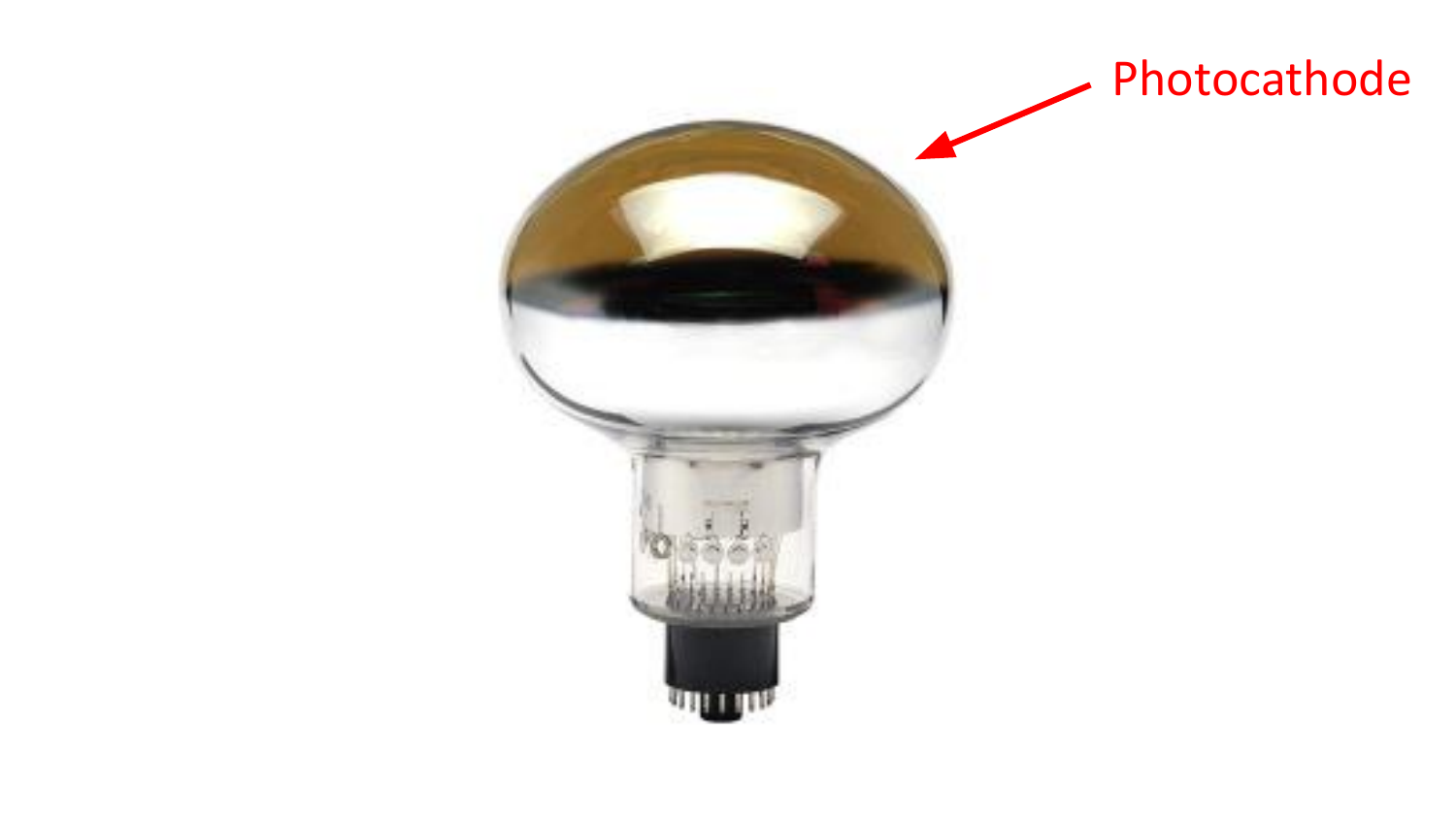} 
\caption{Figure shows a R5912 Hamamatsu PMT~\cite{hama}. The photocathode is shown at the top.}
\label{fig:pmt1}
\end{figure}

The first part of a PMT that usually one observes when looking at one is the photocathode; a glass window at the top of the device (Fig.~\ref{fig:pmt1}). 
This is where electrons are produced. 
Most photocathodes are made of compound semiconductors which consist of alkalic metals with a low work function. 
Hamamatsu Photonics, in their handbook for PMTs, enlists eleven of the most common materials used to fabricate photocathodes~\cite{hama2}.

\begin{figure}[!t]
\centering
\includegraphics[width=7.0cm, height=7.35cm]{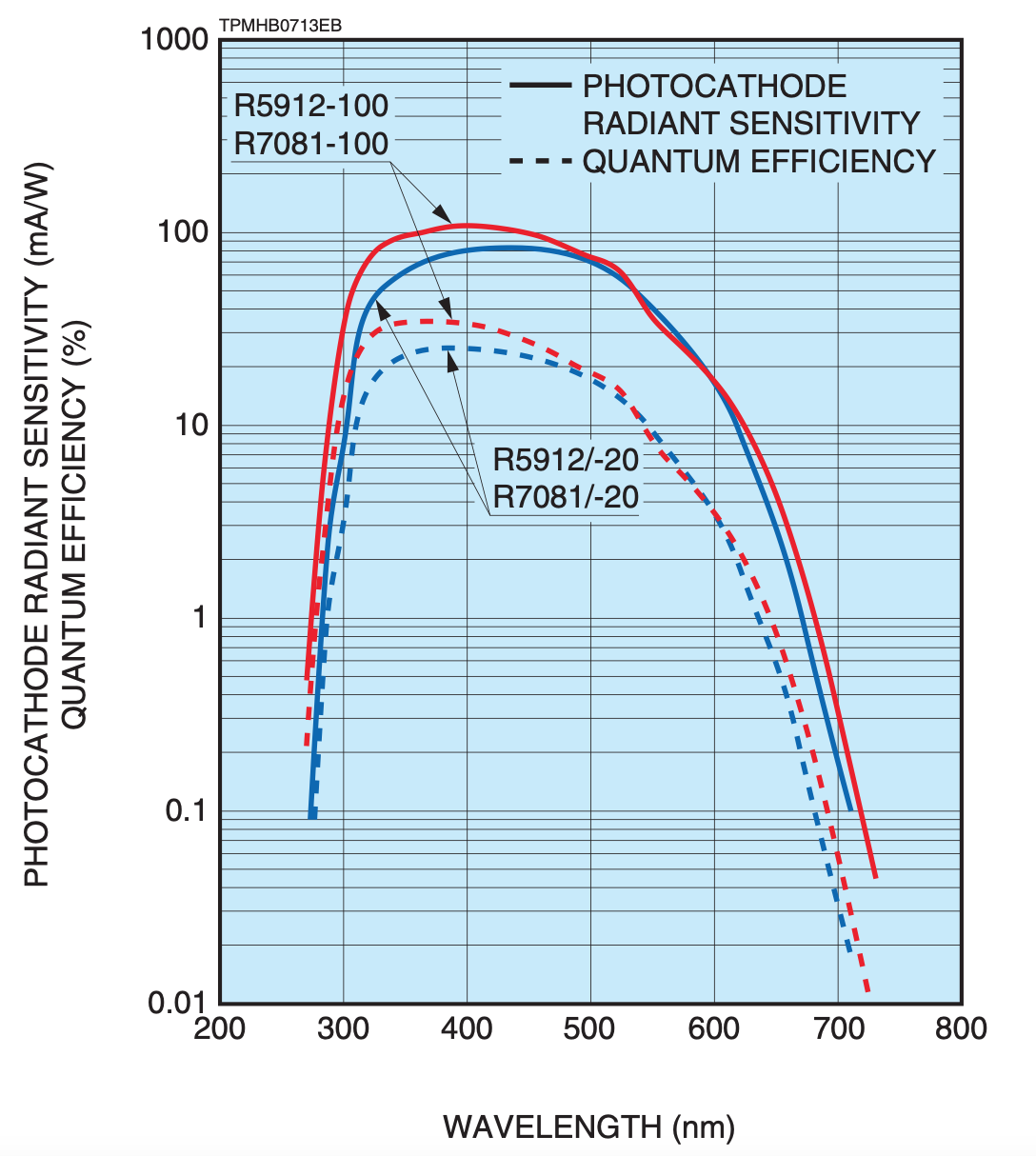} 
\caption{Quantum efficiency as a function of the wavelength of incident light~\cite{qe}.}
\label{fig:qe}
\end{figure}

When a light pulse 
strikes the photocathode there is a certain probability for electrons to be released via external photoelectric effect.  
These electrons are associated with the initial pulse and are commonly known as \emph{photoelectrons} (PEs). 
The probability for this process to happen is called the {quantum efficiency} (QE) and hereafter will be denoted  as $\eta$. It is usually less than $\sim$ 35~\%. 
Of course, $\eta$ depends on the wavelength of the incident light as physics dictates. 
Fig.~\ref{fig:qe} shows the QE as a function of the wavelength for two Hamamatsu PMT models; the R5912 and R7081~\cite{qe}.\footnote{%
Fig.~\ref{fig:qe} also shows the {radiant sensitivity}. This is the photoelectric current generated at the photocathode, divided by the incident flux at a given wavelength. 
We will not concern ourselves with the radiant sensitivity in this document.}
One sees that QE peaks around $\sim$ 380~nm which is in the range of visible blue light. 

After photoconversion, the electrons are accelerated in a series of electrodes, they are focused and finally subjected to a multiplicative dynode structure. 
It is a well-known fact that only a portion of the released PEs survives this secondary process and finally enters the full amplification chain~\cite{hama2}. 
The probability for a PE to reach the first dynode is called {collection efficiency} (CE) and is usually denoted by $\alpha$. 
Note that our notation  agrees with that of the Hamamatsu handbook.
CE usually ranges from approximately (60--90) \% for modern PMTs, but it depends on the high voltage applied. 
Hamamatsu, in their handbook for PMTs, includes a plot of the relative CE as a function of the photocathode to the first dynode voltage~\cite{hama2}.  

\begin{figure}[!t]
\centering
\includegraphics[width=7.5cm, height=6.0cm]{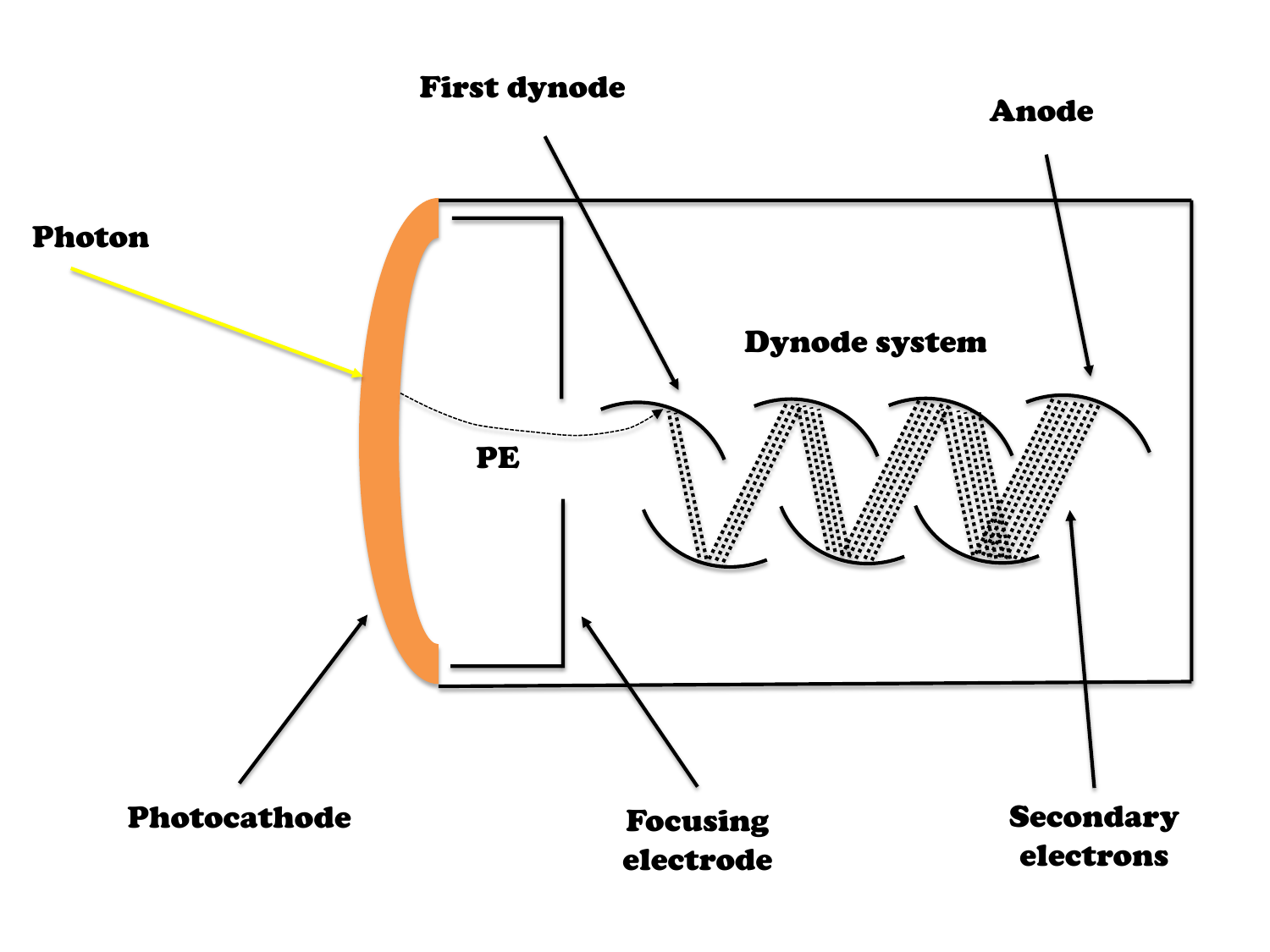} 
\caption{Simple schematic showing the main components of a PMT and the basic principles of its operation.}
\label{fig:pmt2}
\end{figure}

The focused PEs impinge the first dynode and create more electrons by secondary emission. 
This process is then repeated through a series of secondary dynodes until the total charge is collected at the anode. 
At this point the initial charge created in the photocathode has been multiplied by several orders of magnitude. 
The multiplication factor depends on the high voltage applied and it can be as high as $10^8$. 
Fig.~\ref{fig:pmt2} shows the basic steps of PMT operation (photoconversion, focusing, multiplication, etc.) in a simple schematic. 

As mentioned before, the charge amplification in a PMT is a stochastic process. 
This means that the charge produced by a {single photoelectron} (SPE) is not fixed, but it follows from a probability density function (PDF).  
This PDF is commonly known as the \emph{SPE response} (SPER) and it is the most important ingredient in modeling the function of a PMT. 
Knowledge of the charge response allows one to parametrize the behavior of a PMT when illuminated by light pulses.

The mean value of a the SPER is formally called \emph{the gain} $(G)$ and it is of equal importance when it comes to the operation of a PMT. 
It is the proportionality factor that relates charge to PEs. 
To make it simpler, assume that $n$ PEs are fed into the full amplification chain. 
And assume also that each PE produces charge in an independent process.\footnote{A reasonable assumption that holds for every PMT.}
The charge collected at the anode will be on average $Q \simeq n G$.
It is self-evident that the number of PEs creating this event can be approximated to be $n \simeq Q/G$.
 Of course, the approximation only improves as $n$ increases. 
 Knowledge of QE and CE can lead to the number of photons that trigger the PMT and the energy of the particle that created this event. 
Most particle detectors, and especially scintillation calorimeters, are using this recipe to deduce the visible energy of incident particles.  

PMT calibration consists in the determination of the gain and, whenever possible, the SPER. 
There are several reasons that make calibration indispensable. We enlist the following four: 
\begin{enumerate}[i.] 

\item First, one needs to adjust the high voltages of each PMT 
so that they operate at the same gain (nominal high voltages). 
This is a standard recipe and it is known to reduce  energy resolution. 
See, for instance, how this was done for the Double Chooz Inner Veto~\cite{me1}.

\item Second, one needs to monitor the gain of all channels for drifts, or abrupt changes (associated with hysteresis cycles), and correct for those as the experiment is running. 

\item Third, sophisticated algorithms can determine the energy of the incident particle and its vertex if one knows the SPE response functions. 
One can have a look at some of these methods (for large-volume liquid scintillator detectors) in Ref.~\cite{smi1}.

\item Last, in simulating detector responses, using for example Geant4~\cite{geant} or GLG4sim~\cite{glen}, the knowledge of a SPE template plays a very important role.  

\end{enumerate}
In the following pages we present the basic mathematical theory of PMT calibration, in a simple and concise text. 
We describe the standard machinery employed to simulate the behavior of PMTs and extract both gain and SPER. 
These techniques are then applied to a real problem; the calibration of the Hamamatsu R1408 PMT model.

%
%

\section{Light source}
\label{sec:source}

Almost all calibration techniques rely on an inspection of the charge distribution of the PMT when illuminated with faint light pulses containing only a few photons. 
This is the so-called \emph{single photoelectron spectrum}. 
It is customary to prepare the light source in a way that simplifies calculations and makes life easier. 
Most methods employ the use of a poissonian light source:
\begin{align}
P( n; m ) =  \frac{m^n}{n!} \mathrm{e}^{-m}. \label{eq:pois1}
\end{align}
Where $m$ is the mean of the Poisson distribution and $P( n; m )$ is the probability that $n$ photons hit the photocathode. 
The way to make such a source is common and straightforward. 

Assume that the hardware is producing fixed light pulses of $N$ photons. 
Further assume that a strong attenuator (that is a filter) is placed between the source and the PMT. 
The probability for a photon to pass through the filter is equal to $p$. 
In the mathematical language one would say that we deal with $N$ \emph{Bernoulli trials} with the same probability for success. 
The number of photons that make it through the filter is given by a Binomial distribution~\cite{pap}:
\begin{align}
P_p( n | N ) =  \frac{N!}{n!(N-n)!} p^{n} (1-p)^{N-n}. \label{eq:bin}
\end{align}
Let us denote the average of the Binomial distribution by $m = Np$.
Note that this means:
\begin{align}
p =  \frac{m}{N}. 
\end{align}
$P_p( n | N )$ can be rewritten in the following way, by expanding the factorial coefficients and using the definition of the probability $p$. 
\begin{align}
P_p( n | N ) = &  \frac{N(N-1)(N-2) \cdot \cdot \cdot (N-n+1)}{n!} \frac{m^n}{N^n}  \nonumber
		  \\ & \left(1-\frac{m}{N}\right)^{N-n}
\end{align}
Rearranging the terms one gets:
\begin{align}
P_p( n | N ) & =  \frac{N(N-1)(N-2) \cdot \cdot \cdot (N-n+1) }{N^n} \frac{m^n}{n!} \nonumber \\ & \left(1-\frac{m}{N}\right)^{N-n} \nonumber \\
                   & =  \frac{N}{N} \cdot \frac{N-1}{N} \cdot \frac{N-2}{N} \cdot\cdot\cdot \frac{N-n+1}{N}  \cdot \frac{m^n}{n!}  \nonumber \\ & \left(1-\frac{m}{N}\right)^{N-n}. 
\end{align}
Splitting the last term in two we have:
\begin{align}
P_p( n | N ) = &  \frac{N}{N} \cdot \frac{N-1}{N} \cdot \frac{N-2}{N} \cdot\cdot\cdot \frac{N-n+1}{N}  \cdot \frac{m^n}{n!}  \nonumber \\ & \left(1-\frac{m}{N}\right)^{N} \left(1-\frac{m}{N}\right)^{-n}. 
\end{align}
Final part. We take the limit of $N$ going to infinity, while $p$ becomes small so that $m$ remains constant:
\begin{align}
N \rightarrow \infty, \ \ \ p  \rightarrow 0  \ \ \  \textrm{and} \ \ \ Np   \rightarrow m. \nonumber
\end{align}
Then, 
\begin{align}
P_p( n | N )  \rightarrow \frac{m^n}{n!}\mathrm{e}^{-m}. 
\end{align}
Note, and this is important, that the number of initial photons $N$ and the probability $p$ have dropped out from the final formula. 

\begin{figure}[!t]
\centering
\includegraphics[width=7.5cm, height=5.25cm]{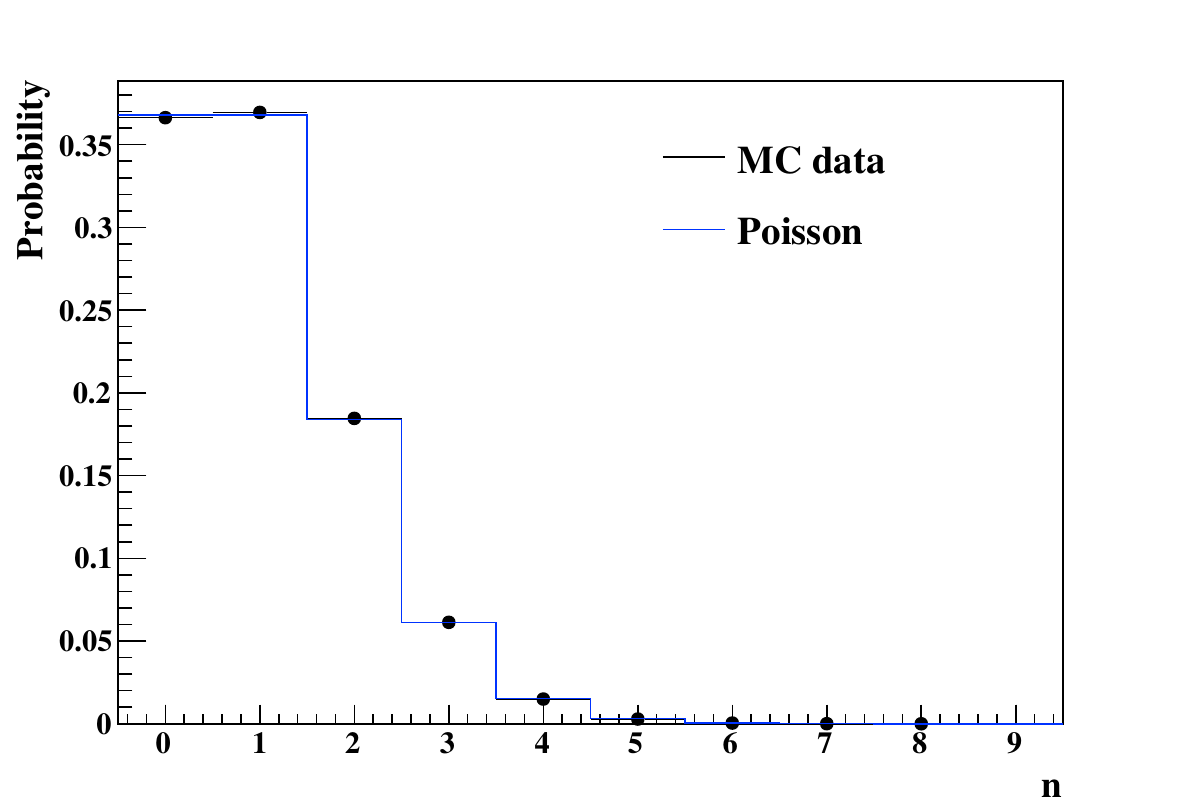} 
\caption{Poisson statistics is a valid approximation when one deals with strong light sources filtered by an equally strong attenuator.}
\label{fig:pois}
\end{figure}

This derivation tells you that a strong light source, blocked by an equally strong attenuator, produces poissonian light pulses. 
To check this result we attempted a toy Monte Carlo (MC) study. 
Several events were generated with 100 photons each. 
For each photon a probability was drawn, form a uniform distribution, and the photon passed the cuts if its probability was smaller that 0.01. 
Plotting these events in a histogram we obtain Fig.~\ref{fig:pois}. 
Black dots show the Monte Carlo data and the blue line the expectation assuming the Poisson distribution of eq.~\eqref{eq:pois1}. 
One sees that generated events and predicted agree very well for these values of $N$ and $p$.  

%
%

\section{Charge response function}
\label{sec:SR}

\subsection{Photoelectron statistics}

The standard technique to describe the charge output of a PMT when illuminated by a poissonian light source was first published in an influential paper written by E. H. Bellamy and collaborators~\cite{Bellamy}. 
In this work,  a sophisticated study was put forward that treats equally both single and multiple photoelectrons alike. 
The first thing that one has to calculate is the number of PEs injected into the multiplicative dynode structure. 
This is not easy, but it can be done along the lines explained in the previous section. 

Assume now that the light pulses that strike the photocathode are described by the probability formula of eq.~\eqref{eq:pois1}. 
As we explained in section~\ref{sec:basic} only a fraction of these photons convert to the photocathode. 
In particular, we expect (on logical grounds) that $\eta m$ PEs will be produced on average. 
Actually, one can prove this theorem rigorously~\cite{pois2} but we will not concern ourselves with this derivation since the final formula we will find, satisfies this condition automatically. 
Photoconversion is a binary process that occurs with probability $\eta$. As explained in previous section it follows from a Binomial distribution. 
The joint probability for $n$ PE production is:
\begin{align}
P_\eta( n ; m ) =  \sum_{N\geq n}^{+\infty} \frac{N!}{n!(N-n)!} \eta^{n} (1-\eta)^{N-n} P( N ; m ). \label{eq:joint}
\end{align}
A few comments are necessary here. 
First, $P( N ; m )$ is the probability to find $N$ photons in the light pulse. 
It is multiplied by the Binomial factors to account for the production of PEs in the photocathode. 
Last, one needs to sum for all possible values of $N$. 
Of course it is obvious that $N$ needs to be larger than $n$ or equal, since  $N$ photons cannot create a larger number of PEs. 

Working the mathematics in eq.~\eqref{eq:joint} we show that:
\begin{align}
P_\eta( n ; m ) & =  \sum_{N \geqslant n}^{+\infty} \frac{N!}{n!(N-n)!} \eta^{n} (1-\eta)^{N-n} P( N ; m ) \nonumber \\
		&  = \sum_{N\geqslant n}^{+\infty} \frac{N!}{n!(N-n)!} \eta^{n} (1-\eta)^{N-n} \frac{m^N}{N!} \mathrm{e}^{-m} \nonumber \\
		& = \frac{\mathrm{e}^{-m} \eta^{n} }{n!} \sum_{N \geqslant n}^{+\infty} \frac{ (1-\eta)^{N-n} m^N  }{(N-n)!} .
\end{align}
Setting
\begin{align}
k = N-n, \nonumber
\end{align}
then one finds:
\begin{align}
P_\eta( n ; m ) & = \frac{\mathrm{e}^{-m} \eta^{n} }{n!} \sum_{k=0}^{+\infty} \frac{ (1-\eta)^{k} m^{k+n}  }{k!}  \nonumber \\
                       & = \frac{\mathrm{e}^{-m} ( \eta m )^{n} }{n!} \sum_{k=0}^{+\infty} \frac{ ( (1-\eta) m ) )^{k}  }{k!}  \nonumber \\
                       & = \frac{\mathrm{e}^{-m} ( \eta m )^{n} }{n!} \mathrm{e}^{ (1-\eta) m } \nonumber  \\
		& =  \mathrm{e}^{-\eta m} \frac{ ( \eta m )^{n} }{n!}.
\end{align}
We see that PEs are produced according to a Poisson distribution of mean value $\eta m$. 
The character of the Poisson distribution is preserved. 

After photoconversion the PEs are filtered according to the collection efficiency as explained in section~\ref{sec:basic}. 
This is again a binary process with probability $\alpha$. 
One can follow the steps described in previous paragraphs only to find that the PEs enter the amplification stage following a Poisson distribution of mean value $\mu = \alpha  \eta m$. 
This is an important result, and it is essentially the starting point of every work on PMT calibration. 
Note, that for this study, we only assumed that we had a strong light source producing pulses of fixed number of photons and coupled with an attenuator. 
We showed that this setup produces PEs according to a Poisson distribution.  
Y. Hu \emph{et al.} have proved that this result holds for a larger class of initial photon statistics~\cite{pois2}.

\subsection{Single photoelectron charge amplification}

In modeling the response of a PMT one is usually forced to postulate a distribution function for the SPE charge deposition. 
The calculation from first principles of this distribution, taking into account all multiplication stages, is very complicated and assumes a very good knowledge of all multiplication parameters~\cite{rade}. 
The SPE response depends on the external conditions such as temperature, magnetic field, etc.
It is characteristic of the PMT and the voltage shared among its dynodes. In what follows it shall be termed as $S(x)$. 

$S(x)\diff x$ gives the probability to collect an amount of charge between $x$ and $x + \diff x$ in the output of the PMT whenever a single PE
is released from the photocathode and focused in the dynode system. 
$S(x)$ is normalized in the usual sense:
\begin{align}
\int_{-\infty}^{+\infty} S(x) \diff x = 1. \label{eq:norm}
\end{align}
It is further assumed that its statistical moments are all well defined. 

In the process of PMT calibration, different types of PDFs are used depending, each time, on the PMT under examination.
For example, the R7081 Hamamatsu PMTs can be adequately described by a gaussian plus an exponential part~\cite{DCID1,DCID2,DCID3}. 
On the other hand, the R1408 Hamamatsu PMT model is effectively parameterized through a gamma distribution plus an exponential part~\cite{me1,me2}. 
We should note that for all the studies conducted for the purposes of this note we were always under the PMT's saturation point. 
In this case, the charge deposition is a linear process with respect to the light intensity and the charge one finally registers is given by the sum of the charges that each PE produces individually.

\subsection{Ideal charge response}

As mentioned before, the PE statistics follows from a Poisson distribution of mean value $\mu$:
\begin{align}
P( n; \mu ) =  \frac{\mu^n}{n!} \mathrm{e}^{-\mu}. \label{eq:pois3}
\end{align}
$\mu$ depends on the initial source, the properties of the attenuator, and the quantum and collection efficiencies jointly. 
It is worth noting that the probability for the light pulse to produce no electrons is different from zero and in fact equals to $P( 0; \mu ) = \mathrm{e}^{-\mu}$.
This means that just by counting the ``zero'' cases, one has direct access to the mean value of the Poisson.
This is an important result that we exploit in later sections of this document.  

Now when no PEs are produced, the PMT (in the absence of electronic noise) will produce charge equal to zero. 
The PDF for the no PE case to produce a charge $x$ will be:
\begin{align}
S_0 (x) =  \updelta (x). 
\end{align}
This is of course 
a delta function centered at zero. 
Meaning that $S_0(x)$ is zero for all $x\neq 0 $ cases, but it peaks at $x=0$ so that it is normalized to one (as it should).  
The total probability for no PE production is given by the product:
\begin{align}
\mathrm{e}^{-\mu} S_0(x)
\end{align}
where $\mathrm{e}^{-\mu}$ is the probability for no PEs to be found in the light pulse. 

Let us now consider the following question: what will be the combined probability for a light pulse, described by eq.~\eqref{eq:pois3}, to produce a single PE that will afterwards deposit an amount of charge between $x$ and $x+\diff x$ ? 
The answer is straightforward from our previous analysis. The total probability is just the product of the two separate probabilities.
\begin{align}
P(1;\mu) S(x) \diff x 
\end{align}
Let us now repeat this question but for the case of a two PE emission. 
So, what will be the combined probability for this light pulse to produce two PEs that will afterwards deposit an amount of charge between $x$ and $x + \diff x$ ? 
The answer is not so obvious.

The total probability for the first PE to deposit a charge between $y$ and $y$ + $\diff y$, 
and the second PE charge between $z$ and $z$ + $\diff z$ is given by the product of the two separate  probabilities since these processes are fully independent: 
\begin{align}
S(y)\diff y  S(z)\diff z.
\end{align}
Denoting as $x = y+z$ the total charge registered, this probability can by rewritten in the form: 
\begin{align}
S(x-z) \diff x  S(z) \diff z,
\end{align}
by means of a simple substitution. 
To obtain the probability for a total charge deposition between $x$ and $x$ + $\diff x$ one has to integrate over all $z$ charges. 
\begin{align}
\left(  \int_{-\infty}^{+\infty} S(x-z) \ S(z)\diff z  \right) \diff x 
\end{align}
The expression in the parenthesis is the convolution of $S(x)$ with itself,  $(S*S)(x)$, 
and as a matter of fact it is a trivial mathematical result that in such cases the total probability distribution is given by the convolution of the two initial PDFs~\cite{pap}. 
Thus, two PEs produce a total charge $x$ according to the probability distribution $S_2(x)$ where:
\begin{align}
S_2(x) = (S * S )(x).
\end{align}

We are now in position to answer our second question, and the answer is a simple generalization of our first result:
\begin{align}
P(2;\mu) S_2(x) \diff x.
\end{align}
Of course, similar arguments hold for the case of three PEs, four PEs, and so forth. 
To generalize these formulae, we denote $S_1(x)=S(x)$, so that the $n$PE probability can be written as: 
\begin{align}
P(n;\mu) S_n(x) \diff x, 
\end{align}
where $S_n(x)$ for $n \geqslant 2$ is the $n$-times convolution of $S(x)$. 
We reco\-gnize that the total probability for a light pulse to produce charge between $x$ and $x+\diff x$ is given by: 
\begin{align}
\left(  \sum_{n=0}^{+\infty} P( n; \mu ) S_n(x)  \right) \diff x,
\label{eq:Sid}
\end{align}
where eq.~\eqref{eq:Sid} is the sum of the individual probabilities for any given number of PEs.
We repeat that $S_0(x)=\updelta(x)$, $S_1(x)=S(x)$ and $S_n(x)$ for $n \geqslant 2$ is the $n$-times convolution of $S(x)$.

\subsection{Including background noise}

The PDF for the PMT charge output in response to a poissonian light source is termed $S_{ID}(x)$. 
The form of $S_{ID}(x)$ was worked out in the previous subsection and it is repeated here: 
\begin{align}
S_{ID}(x) = \sum_{n=0}^{+\infty} P( n; \mu ) S_n(x). 
\label{eq:Sid2}
\end{align}
We choose to call this PDF $S_{ID}(x)$ because it is ideal in the sense that it cannot account for any external charge fluctuations as those produced by electronic noise.
Indeed, never in our previous analysis did we mention about the PMT pedestal and the charge smearing that's responsible for. 

To incorporate the background charge in this formalism, a shifted gaussian over a mean value $Q_0$ is used. 
\begin{align}
B(x) = \frac{1}{\sqrt{2\pi}\sigma_0} \mathrm{e}^{ - \frac{( x - Q_0 )^2}{2\sigma_0^2}}
\end{align}
$\sigma_0$ describes the charge smearing induced by the electronic noise.
For the same reasons we needed to convolute the signal distribution $S(x)$ with itself to obtain the two PE $S_2(x)$ distribution, 
here we need to convolute $S_{ID}(x)$ and $B(x)$ distributions to produce the output signal PDF $S_{R}(x)$. 
\begin{align}
S_R(x) = &     (S_{ID}*B)(x) \nonumber  \\
            = & \sum_{n=0}^{+\infty} P( n; \mu ) ( S_n * B )(x)
\label{eq:Sr0}
\end{align}
Here ``$R$'' stands for real or realistic. 
In general,  our job here is to derive $S_R(x)$ in a closed or, at least, in an approximate but still useful form.

%

At first this might seem like an extraordinary enterprise but a valid approximation can be easily derived if the mean value of the Poisson distribution ($\mu$) is low enough. 
In that case, if $\mu$ is small, $S_R(x)$ can be written as: 
\begin{align}
S_{R}(x) \simeq \sum_{n=0}^{k} P( n; \mu ) (S_n*B)(x), 
\label{eq:Sr}
\end{align}
where $k$ is an integer chosen such that the $ P( n; \mu )  (S_n*B)(x)$ terms with $n > k $ are negligible owing to the dumping of the Poisson factors. 
When $\mu= 1$,  $k$ can be chosen to be equal to five, while for $\mu= 2$ the first seven terms of eq.~\eqref{eq:Sr} are enough to give a precision better than 1~\%.

The aforementioned recipe was first presented in Ref.~\cite{Bellamy}.  
In that paper, the authors give an exact formula for the $(S_n*B)(x)$ distributions when the SPE response function $S(x)$ is given by a symmetric gaussian. 
$S_R(x)$ can then be used to fit real data taken while the PMT is illuminated with low-intensity light pulses and the parameters of $P(n;\mu)$, $B(x)$ and $S(x)$ can be extracted. 
Note that the PMT gain $G$ is given by:  \enlargethispage{\baselineskip}
\begin{align}
G =  \int_{-\infty}^{+\infty} x S(x)  \diff x,
\end{align}
which is the mean value of $S(x)$.

An important improvement of the original Bellamy work has been put forward by O. Ju. Smirnov and collaborators in Ref.~\cite{Smirnov}. 
In particular, the model they proposed is described by a combination of a truncated gaussian plus an exponential distribution:
\begin{align}
S(x) =    \left( \ w \alpha \mathrm{e}^{-\alpha x } + \frac{(1-w)}{g_N} \frac{1}{\sqrt{2\pi}\sigma} \mathrm{e}^{ - \frac{( x - Q )^2}{2\sigma^2}} \ \right) H( x ),
\label{eq:smi}
\end{align}
where
\begin{align}
g_N =  \frac{1}{2} \text{erfc} \left( -\frac{Q}{\sqrt{2}\sigma } \right). 
\end{align}
erfc($z$) is the complementary error function~\cite{error} and the Heaviside $H(x)$ step function is necessary because negative charges are not physical. $g_N$ is defined so that $S(x)$ is normalized to one. 
The key elements of the Smirnov model can be summarized in the following points:
\begin{enumerate}[i.]
\item The realization that the exponential term that models PEs that miss the first amplification stage is part of the signal (it adds to the gain) and,
\item The use of a properly normalized, truncated gaussian to describe the full amplification chain, avoiding thus the prediction of negative charges in the underlying formulae. 
\end{enumerate}
None of these points is entirely trivial. 
Nonetheless, an $S_R(x)$ model based on eq.~\eqref{eq:smi} is quite difficult to derive in a closed form.
One usually resorts to some sort of an approximation, or solve the convolution integrals numerically.\footnote{%
Although, note that an exact formula of the Smirnov model has been given in Ref.~\cite{ana} with only the minimum amount of approximations. 
It is based on special functions and the final solution is a bit complicated.}  

There are several methods that can be used to measure the gain of a PMT. 
Some require the knowledge of an analytical $S(x)$ model, and others can extract $G$ in a model independent manner. 
Each of them share its own advantages and disadvantages. 
In the following pages, we present the main techniques 
for PMT calibration.  
But first, it is instructive to pause a bit and calculate the mean value and standard deviation of $S_R(x)$. 
If not for no other reason, these formulae are needed for the \emph{occupancy approach} described in section~\ref{sec:occ}.

%
%

\section{Mean value and variance of $S_R(x)$}
\label{sec:mean}
%

%

In order to compute the mean value ($Q_R$) and standard deviation ($\sigma_R$) of the charge response $S_R(x)$, we write down the formulae:
\begin{align}
\sum_{n=0}^{+\infty} P( n; \mu ) & = 1, \\
\sum_{n=0}^{+\infty} n P( n; \mu ) & = \mu, \\
\sum_{n=0}^{+\infty} n^2 P( n; \mu ) & = \mu ( \mu + 1 ). 
\end{align} 
The first equation stems from probability conservation and it is very easy to derive.
The other two can be proved by a shifting of the summing parameter $n$. 

Let us now denote by $Q_0$ and $\sigma_0$ the mean value and standard deviation of $B(x)$ respectively.
Additionally, we use the symbols $G$ and $\sigma_G$ for the mean value and standard deviation of  $S(x)$. 
Of course, $G$ is the gain. 
We also note  the formulae for the mean value (E) and variance (Var) of $( S_n * B )(x)$ that are deduced from the properties of the convolution:
\begin{align}
\text{E}[( S_n * B )(x)]  \ = \ & Q_0 + n  G, \\
\text{Var} [ ( S_n * B )(x) ] \ = \ & \sigma_0^2 + n\sigma_G^2. 
\end{align} 
These equations are calculated in detail in the \ref{app:a}. 
We simplify our notation by setting $S_R^{(n)}(x)=( S_n * B )(x)$. 
Using these equations, the mean value $Q_R$ becomes:
\begin{align}
Q_R \ = \ & \text{ E }[ S_R (x)]  \nonumber \\
 \ = \ & \sum_{n=0}^{+\infty} P( n; \mu ) \int_{-\infty}^{+\infty} x S_R^{(n)}(x) \ \diff x \nonumber \\
  \ = \ & \sum_{n=0}^{+\infty} P( n; \mu ) ( Q_0 + n G ) \nonumber \\
  \ = \ & Q_0 + \mu G.   \label{eq:G}
\end{align} 
On the other hand, to calculate the variance $\sigma_R^2$ we first find the integral:
\begin{align}
\int_{-\infty}^{+\infty} x^2 S_R(x) \ \diff x \ = & \  \sum_{n=0}^{+\infty} P( n; \mu )   \int_{-\infty}^{+\infty} x^2 S_R^{(n)}(x) \ \diff x \nonumber \\
\ = & \ \sum_{n=0}^{+\infty} P( n; \mu ) ( \sigma_0^2 + n\sigma_G^2 \nonumber \\ & + ( Q_0 +n G )^2 ) \nonumber \\
\ = & \ \sum_{n=0}^{+\infty} P( n; \mu ) ( \sigma_0^2 + n\sigma_G^2 +  Q_0^2  \nonumber \\ & + n^2 G^2  + 2 n Q_0 G ) \nonumber \\
\ = & \ \sigma_0^2 + \mu\sigma_G^2 + Q_0^2 + \mu ( \mu + 1 ) G^2 \nonumber \\ & + 2 \mu Q_0 G \nonumber \\
\ = & \ \sigma_0^2 + \mu(\sigma_G^2 +  G^2 ) + ( Q_0 + \mu G )^2. \label{eq:sG}
\end{align}
Where in the forth line we made use of the identities provided in the beginning of this section. 
The variance now becomes:
\begin{align}
\sigma_R^2 \ = \ & \text{Var}[ S_R(x)] \nonumber \\
\ = \ & \int_{-\infty}^{+\infty} x^2 S_R(x)  \diff x - Q_R^2 \nonumber \\
\ = \ & \sigma_0^2 + \mu( \sigma_G^2 + G^2 ).  \label{eq:sG}
\end{align}
Both eqs.~\eqref{eq:G} and \eqref{eq:sG} will turn out to be useful when we discuss the task of gain determination in later sections of this document. 


%
%

\section{Numerical methods}
\label{sec:num}

In order to calculate $S_R(x)$ it is obvious from eq.~\eqref{eq:Sr} that one needs to find the $(S_n * B)(x)$ distributions.
At first sight, this might seem like an easy job, but a careful look of the formulae forces one to realize that only in the simplest models one can derive $(S_n * B)(x)$ in a closed form. 
Even without the final convolution with $B(x)$, in several cases it is even difficult to calculate the $n$PE response functions $S_n(x)$. 
Take for instance Smirnov's model of eq.~\eqref{eq:smi}. 
For $n=2$ one has three terms to compute, while for $n=3$ one has four different terms in $S_n(x)$.
And this number only increases as $n$ becomes larger !
In the original Smirnov \emph{et al.} paper a simple formula was given where all $S_n(x)$ distributions above $n=1$ were modeled by symmetric gaussians.\footnote{%
Note that the original Bellamy model is also an approximation~\cite{Bellamy}. }

For us the battle is won !
Instead of relying on crude approximations that can invalidate the results, one can rather calculate $S_R(x)$ numerically. 
If the precision of the estimation is good enough, one can have excellent results without worrying whether the underline assumptions are violated or not. 
There are two commonly known numerical methods in the literature:
\begin{enumerate}[i.]
\item  A method that computes the convolution integrals numerically, using a recursive algorithm, 
for the first PE peaks and approximates higher order corrections with symmetric gaussians. 
\item And another technique, based on the Discrete Fourier Transform (DFT), that calculates the charge response function $S_R(x)$ numerically for all orders in the Poisson mean $\mu$.
\end{enumerate}
Both procedures give consistent results when carefully applied. 
In this section we describe in some detail the essential points of these methods. 
 
\subsection{Numerical Integration}

Solving the convolution integrals numerically is, of course, the obvious thing to do.
Nonetheless, to  our knowledge, this recipe was first proposed by Smirnov \emph{et al.} in Ref.~\cite{Smirnov}. 
It is perhaps interesting to wonder how many novelties were introduced in this publication~!
The range 
was digitized in a grid of $M$ points; $x_i$ with $i = 1 \rightarrow M$. 
$x_i$ were arranged from smaller to larger in ascending order. 
Now the width of this grid, $\Delta x = x_{i+1} - x_i$, can be chosen to coincide with the binning of the SPE histogram or it can be even made finer to increase precision. 
Running on this grid, the convolution integral:
\begin{align}
S_n(x) = \int_{-\infty}^{+\infty} S_{n-1}(z) S(x-z) \diff z,
\end{align}
can be approximated by:
\begin{align}
S_n(x_i) \simeq \sum_{j=1}^{i} S_{n-1}(x_j) S(x_i-x_j) \Delta x, \label{eq:sum1}
\end{align}
where $x_j$ is a running variable ranging from $x_1$ to $x_i$. 
Note that for $j>i$, $(x_i-x_j)$ becomes negative and $S(x_i-x_j)$ vanishes along the lines described in section~\ref{sec:SR}.
Likewise the sum of eq.~\eqref{eq:sum1} stops at $j=i$.
In this way one can calculate recursively $S_n(x_i)$ for any value of $n$. 
Meaning that you first derive $S_2(x_i)$, then $S_3(x_i)$, ... and so on.

For a sufficiently large a value of $n$, it is a well-known mathematical theorem 
that the $n$-times convolution will approach a gaussian probability density~\cite{pap}.
In this case the mean of the gaussian will be:
\begin{align}
Q_n = n G,
\end{align}
and the standard deviation: 
\begin{align}
\sigma_n = \sqrt{n} \sigma_G.
\end{align}
The formula for $S_n(x)$ becomes:
\begin{align}
S_n(x_i) \simeq \frac{1}{\sqrt{2\pi}\sigma_n} \mathrm{e}^{ - \frac{( x_i - Q_n )^2}{2\sigma_n^2}}. 
\end{align}
It has been shown by Saldanha \emph{et al.} that for the Smirnov model, the $S_n(x)$ distributions become symmetric gaussians to an excellent precision for $n\geqslant 10$~\cite{salda}. 

Armed with these formulae, the ideal response function of eq.~\eqref{eq:Sr} can be written:
\begin{align}
S_{ID}(x_i) \simeq \sum_{n=0}^{k} P( n; \mu ) S_n(x_i).
\end{align}
Now the final convolution with $B(x)$ can be performed along similar lines, but with an important simplification. 
Namely, we note that for $x<Q_0-5\sigma_0$ or $x>Q_0+5\sigma_0$ the gaussian of $B(x)$ is expected to vanish. 
Finding the bin where the maximum of $B(x)$ lies one can isolate a symmetric range of bins $M_1 - M_2$, around the maximum, that covers the region between $Q_0-5\sigma_0 < x < Q_0+5\sigma_0$. 
For the $M_1$-th bin we have:
\begin{align}
x_0 + M_1 \Delta x \simeq Q_0-5\sigma_0,
\end{align}
and likewise for $M_2$:
\begin{align}
x_0 + M_2 \Delta x \simeq Q_0+5\sigma_0.
\end{align}
$x_0$ is the lower edge of the first bin. 
Limiting the summation between $M_1$ and $M_2$ one can save a lot of computational time. 
Like this the final formula for $S_{R}(x)$ reads:
\begin{align}
S_{R}(x_i) \simeq \sum_{j=M_1}^{M_2} B(x_j) S_{ID}(x_i-x_j) \Delta x. \label{eq:sum2}
\end{align}
We repeat that the numerical integration will yield rigorous results when carefully applied. 
Special attention requires the approximation with perfect gaussians. 
For instance, the DarkSide collaboration calculated the first two PE peaks numerically, while for $n \geqslant 3$ the gaussian approximation was employed~\cite{dark}. 
This was adequate for the needs of DarkSide. 
This method though has, in our own (and subjective) view two main disadvantages: 
\begin{enumerate}[i.]
\item  First, it appears to be quite tedious to be implemented efficiently in an analysis software.
Especially the part of the code where the recursive algorithm for the calculation of $S_n(x_i)$ is introduced can be tricky, and
\item Second, it is known that it requires a lot of computational time to calculate all the for loops needed for the derivation of $S_R(x)$~\cite{me3}.
\end{enumerate}
To bypass both these shortcomings, an alternative technique has been proposed in Ref.~\cite{me2} that we now turn our attention to. 

\subsection{Discrete Fourier Transform approach to $S_R(x)$}

To start an exposition of this numerical method, we first take the Fourier Transform (FT) of $S_R(x)$ denoted here as $\tilde S_R(k)$. 
It is a well-known result of Fourier analysis that the FT of a convolution is equal to the product of the FTs of each individual function.\footnote{%
See for instance Chapter 15 of Ref.~\cite{Arfken}. }
With these considerations, the FT $\tilde S_R(k)$ can be written as: 
\begin{align}
\tilde S_R(k) = \sum_{n=0}^{+\infty} P( n; \mu ) \tilde S^n(k)  \tilde B(k). 
\label{eq:ft}
\end{align}
$\tilde S^n(k)$ is the $n$-th power of $\tilde S(k)$, the FT of the SPE response function $S(x)$, and $ \tilde B(k)$ is the FT of the pedestal noise. 
Owing to the mathematical form of the Poisson $P( n; \mu )$ factors, eq.~\eqref{eq:pois3}, the series of eq.~\eqref{eq:ft} can be formally summed and the result is: 
\begin{align}
\tilde S_R(k) = \mathrm{e}^{-\mu} \tilde B(k) \mathrm{e}^{ \mu \tilde S(k) }. 
\label{eq:master}
\end{align}
Of course, it is obvious that the Inverse FT of eq.~\eqref{eq:master} gives $S_R(x)$. 

If one could invert the formula of eq.~\eqref{eq:master}, a closed form for $S_R(x)$ could be obtained. 
In practice this is an impossibility. In several cases, even a straightforward calculation of $\tilde S(k)$ is not at all trivial. 
To bypass these obstacles, it was proposed to evaluate FTs (and their inverse) numerically using Discrete Fourier Transform (DFT). 
If the steps chosen in DFT are small enough, a good estimation of $S_R(x)$ can be achieved. 
Additionally, these operations are fast by virtue of the many excellent Fast Fourier Transform (FFT) algorithms currently available. 

In practice, we employ the well-known C library FFTW~\cite{fftw} for the evaluation of the real ($\Re$) and imaginary ($\Im$) parts of both $\tilde S(k)$ and $\tilde B(k)$.
Eq.~\eqref{eq:master} can then be written as: 
\begin{align}
\tilde S_R(k) = \mathrm{e}^{-\mu} | \tilde B(k) | \mathrm{e}^{  \mu\Re[  \tilde S(k) ] }   \mathrm{e}^{i ( \phi_{\tilde B} + \mu \Im[\tilde S(k) ] )}, 
\label{eq:master2}
\end{align} 
where $\Re[ \tilde S(k) ]$ and $\Im[ \tilde S(k) ]$ are the real and imaginary parts of $\tilde S(k)$, and $| \tilde B(k) |$, $\phi_{\tilde B}$ are the magnitude and argument of $\tilde B(k)$ respectively. 
The complex parts of $\tilde S_R(k)$ are given by:
\begin{align}
\Re[\tilde S_R(k) ] & = \mathrm{e}^{-\mu} | \tilde B(k) | \mathrm{e}^{ \mu \Re[ \tilde S(k) ]  }  \cos( \phi_{\tilde B }+ \mu \Im[\tilde S(k) ] ) \ \text{and}, \\
\Im[\tilde S_R(k) ] & = \mathrm{e}^{-\mu} | \tilde B(k) | \mathrm{e}^{  \mu \Re[ \tilde S(k) ]  }  \sin( \phi_{\tilde B } + \mu \Im[\tilde S(k) ] ). 
\label{eq:parts}
\end{align} 
Inserting $\Re [ \tilde S_R(k) ]$ and  $\Im [ \tilde S_R(k) ]$ in the Inverse DFT (IDFT) implemented in the FFTW library one can then retrieve the values of $S_R(x)$ numerically. 

For all the analyses presented in this document, a step equal to the size of the histogram binning was used for the DFTs and IDFT. Here, this was sufficient to reach the required accuracy level. 
A  \texttt{C++/ROOT} based software~\cite{root}  that implements the DFT approach and evaluates $S_R(x)$ is available in \cite{git}, a public \texttt{git-hub} repository.
The code is simple, with several examples, and can be used to plot $S_R(x)$ given that the parameters of $S(x)$ are all known. 
It is quite efficient in the sense that it only needs fractions of  a second to draw a single distribution.  
Last, we should emphasize that, in contrast to eq.~\eqref{eq:Sr} the DFT method is able to derive $S_R(x)$ for any value of $\mu$ since eq.~\eqref{eq:master} represents the full sum of the $S_R(x)$ series. 
Very recently, the JUNO collaboration has employed the DFT approach to determine the gains of the twenty inch PMTs that monitor its target material~\cite{perf}.

%
%

\section{Single photoelectron model}
\label{sec:spe}

Before we discuss the intricate details of gain determination, it is perhaps advisable to plot $S_R(x)$ and note its basic characteristics. 
To do so one has to select a model for the SPE response function. 
The obvious choice would be to pick Smirnov's model because of its simplicity, and because it is rich in scope and far richer in its range of applicability. 
Indeed most PMTs have symmetric SPERs that can be effectively parametrized by a truncated gaussian. 
Nonetheless, in this publication we analyzed a series of large statistics data samples taken with a Hamamatsu eight inch R1408 PMT.  
Now it has been shown that the R1408 are best described by a combination of an exponential plus a gamma distribution~\cite{me1,me2}.
And since we would like to compare Monte Carlo (MC) with real data we will stick to this later model of SPER throughout this document.   

In general, the R1408 PMTs are difficult to calibrate mainly for two reasons: 
\begin{enumerate}[i.]
\item The various PE peaks are entangled due to the significant dispersion of the charge; 
a fact that complicates the extraction of the gain via conventional methods\footnote{%
The reader might have a look at Fig.~3 of Ref.~\cite{Bellamy}, where the 2nd PE peak is visible even with naked eye, for an example of a case where there is little charge dispersion.} and, 
\item The SPE response function $S(x)$ cannot be parametrized by a gaussian (truncated or not) and the final model for $S_R(x)$ cannot be solved analytically.    
\end{enumerate}
Fortunately, it has been shown that PMT model R1408 can be treated if one relies on the probability density function~\cite{me2}:
\begin{align}
S(x) =   & \ w \alpha \mathrm{e}^{-\alpha x } H(x)   \nonumber \\ & +  ( 1 - w ) \lambda (1+ \theta ) \frac{[ \lambda (1+ \theta ) x ]^\theta }{\Gamma (1+\theta) } \mathrm{e}^{ - \lambda (1+ \theta ) x } H( x ).
\label{eq:polya}
\end{align}
The purely exponential term of eq.~\eqref{eq:polya} (the first term) is absolutely necessary to describe the amplification of PEs that miss the first dynode or photons that convert directly in the first dynode. 
The slope of the exponential is $\alpha$ and the pre-factor $w$ parametrizes the probability for this to happen. 
A better discussion of this process, with further experimental tests, can be found in Ref.~\cite{Smirnov}. 

\begin{figure}[!t]
\centering
\includegraphics[width=7.5cm, height=5.25cm]{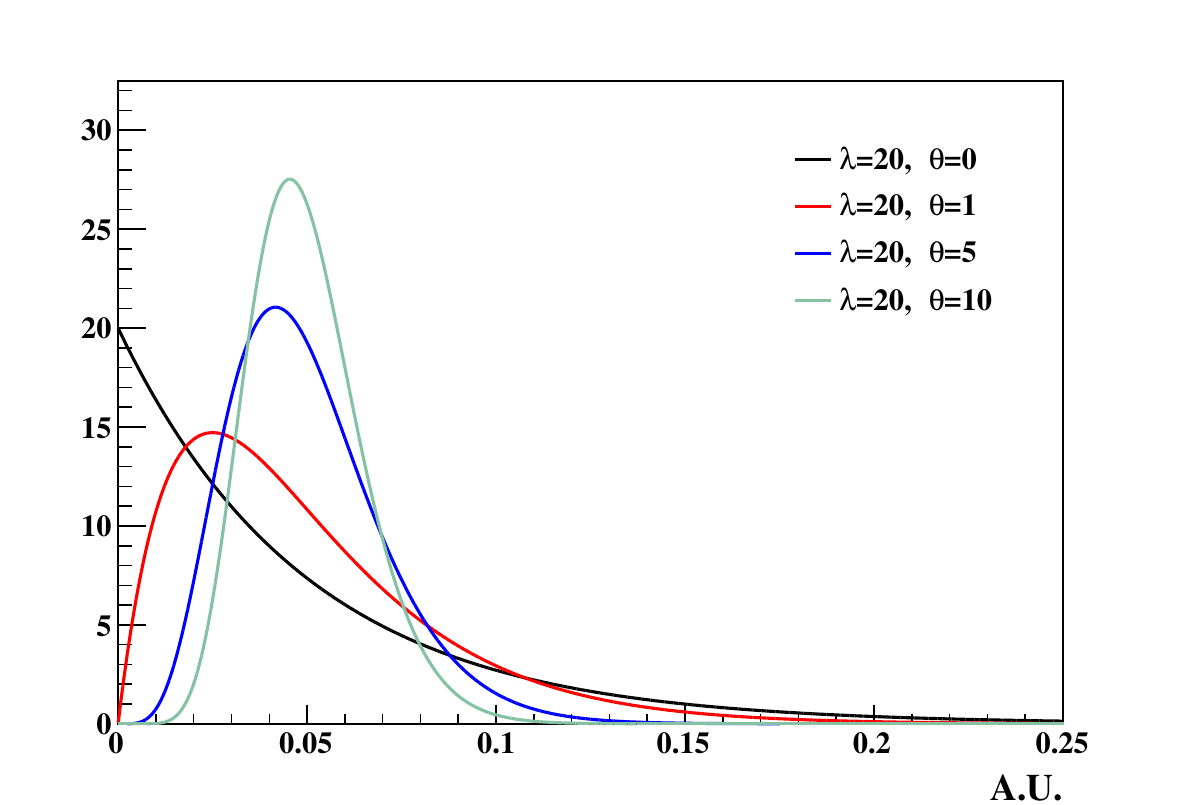} 
\caption{Gamma distribution for $\lambda=20$ and some values of $\theta$. 
As $\theta$ increases the standard deviation of the distribution decreases and its shape becomes more narrow. }
\label{fig:polya}
\end{figure}

The second term is a gamma distribution and it describes the full dynode amplification chain. 
This distribution has been used in the past in connection with the R1408 PMTs calibration~\cite{me1}. 
Note that in this guise, the gamma distribution is also known in the physics vocabulary as \emph{the Polya distribution}. 
Physicists understood its utility in the 60's and 70's in relation to the then newly discovered multiwire gas chambers~\cite{wire}.
When $\theta=0$ the gamma distribution degenerates into a pure exponential. 
On the other hand, when $\theta$ increases the standard deviation decreases and the distribution acquires a more narrow and peaked shape. 
Figure~\ref{fig:polya} shows a few curves for some random values of $\lambda$ and $\theta$. 
In this respect, this model is rather general and it can be used to describe a vast range of PMTs. 

The final model of $S_R(x)$ was solved numerically using the DFT machinery developed in previous section,
since the mathematics involved in the calculations of the $S_n(x)$ and $(S_n*B)(x)$ convolutions cannot be solved analytically. 
We should repeat that we could have used the numerical integration method to derive $S_R(x)$, 
but we decided to stick to the DFT approach mainly because: (a) it appears to be simpler and (b) it is much faster. 
In general, in this article we will not use the numerical integration to analyze data. 
Of course, and we would have gotten the same results with DFT, albeit with a larger amount of time. 

\begin{figure}[!t]
\centering
\includegraphics[width=7.5cm, height=5.25cm]{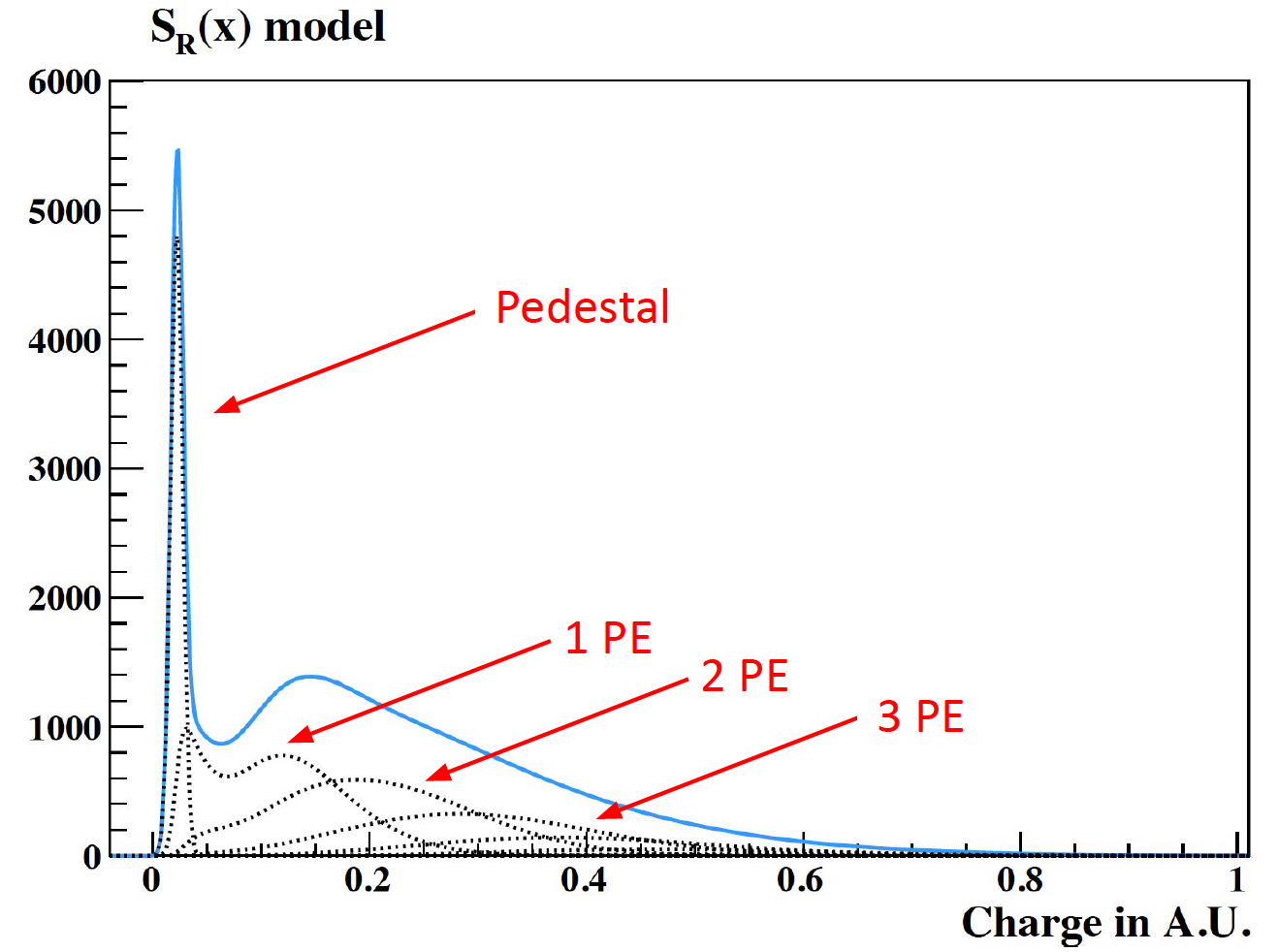} 
\caption{Plot of the $S_R(x)$ distribution in the azure line. The black dotted lines show the contributions of the various PEs.}
\label{fig:plot}
\end{figure}

Now to plot $S_R(x)$ one has to chose the parameter values of $B(x)$ and $S(x)$. 
We decided to use some realistic numbers, and picked those from the first row of table~1 in Ref.~\cite{me2}. 
In particular, for the background noise we have:
\begin{align}
Q_0 = 0.02199, \  \sigma_0 = 0.00562.
\end{align}
The exponential  slope was fixed to:
\begin{align}
\alpha = 23,
\end{align}
and the $w$ pre-factor was set to:
\begin{align}
w = 0.41.
\end{align}
Finally, the gamma distribution parameters were:
\begin{align}
\lambda = 7.79, \ \theta = 5.1.
\end{align}
Fig.~\ref{fig:plot} shows the plot of $S_R(x)$ for some random overall normalization and for the poissonian mean equal to two, $\mu=2$.
$S_R(x)$ is shown in the azure line, and the several PE contributions in black dotted lines. 
A few remarks are necessary here. The first thing that one observes looking at this figure is a pronounced peak close to zero. 
This corresponds to the $\mathrm{e}^{-\mu}B(x)$ term in eq.~\ref{eq:Sr0} (the first term) and it is commonly known as \emph{the pedestal}. 
These are the cases when no PEs were observed by the PMT. 
Second, one does not see a sharp SPE peak in this spectrum. Quite to the contrary the various PEs are seriously entangled. 
We can only expect that this feature can complicate the estimation of the gain via conventional methods.
Third, and last one, armed with these concepts and formulae were are in a position to tackle the simplest and most straightforward approach to gain determination. 

%
%

\section{Mean value and variance of $S(x)$}
\label{spemean}

But before we present the complex details of gain determination, it is perhaps advisable to pause a bit and compute the mean ($G$) and standard deviation ($\sigma_G$) of the gamma function model $S(x)$. 
The mathematics involved are simple and beautiful and, if not for no other reason, these formulae are needed in later sections of this report. 
Now, $S(x)$ can be written as:
\begin{align}
S(x) =  w f(x) + (1-w) g(x),
\end{align}
where 
\begin{align}
f(x) = \alpha \mathrm{e}^{-\alpha x} H(x), 
\end{align}
and 
\begin{align}
g(x) = \lambda (1+ \theta ) \frac{[ \lambda (1+ \theta ) x ]^\theta }{\Gamma (1+\theta) } \mathrm{e}^{ - \lambda (1+ \theta ) x }  H(x).
\end{align}
To calculate $G$ and $\sigma_G$ one needs the mean value and variance of both $f(x)$ and $g(x)$.
It has to be noted that $f(x)$ is a special case of $g(x)$ when $\theta = 0$. 
This means that solving the equations for the mean ($Q_g$) and standard deviation ($\sigma_g$) of $g(x)$ one can derive those of $f(x)$ by means of a simple substitution. 
Naturally, we first treat the gamma distribution. 

The mean value of $g(x)$ is given by the formula:
\begin{align}
Q_g & = \int_{-\infty}^{+\infty} x \ g(x) \diff x \nonumber \\
& = \frac{[ \lambda (1+ \theta ) ]^{1+\theta} }{\Gamma (1+\theta) } \int_{0}^{+\infty} x^{1+\theta} \ \mathrm{e}^{ - \lambda (1+ \theta ) x }  \diff x.
\end{align}
With the substitution:
\begin{align}
u =  \lambda (1+ \theta ) x,
\end{align}
$Q_g$ can be written as:
\begin{align}
Q_g = \frac{1}{\lambda ( 1+\theta ) } \frac{1}{\Gamma (1+\theta)} \int_0^{+\infty} u^{1+\theta} \ \mathrm{e}^{-u} \diff u. \label{eq:qg1}
\end{align}
The integral of eq.~\eqref{eq:qg1} can be solved using the definition of the gamma function~\cite{error}:
\begin{align}
\Gamma (z) = \int_0^{+\infty} x^{z-1} \ \mathrm{e}^{-x} \diff x, \ \label{eq:propg1}
\end{align}
and like this, $Q_g$ becomes:
\begin{align}
Q_g = \frac{1}{\lambda ( 1+\theta ) } \frac{1}{\Gamma (1+\theta)} \Gamma ( 2+\theta ). 
\end{align}
This equation can be further simplified by the property of the gamma function~\cite{error}:
\begin{align}
\Gamma (z+1) = z \Gamma (z), \ \label{eq:propg2}
\end{align}
and finally one gets:
\begin{align}
Q_g = \frac{1}{\lambda}. \label{eq:qg5}
\end{align}

The calculation of the standard deviation of $g(x)$ proceeds in a similar way. 
Actually, to compute $\sigma_G$ one needs (first) to find  the integral:
\begin{align}
\int_{-\infty}^{+\infty} x^2 \ g(x) \diff x & = \frac{[ \lambda (1+ \theta ) ]^{1+\theta} }{\Gamma (1+\theta) } \int_0^{+\infty} x^{2+\theta} \ \mathrm{e}^{ - \lambda (1+ \theta ) x }  \diff x \nonumber \\
& = \frac{1}{\lambda^2 } \frac{1}{(1+\theta)^2 } \frac{1}{\Gamma (1+\theta)} \int_0^{+\infty} u^{2+\theta} \ \mathrm{e}^{ - u }  \diff u.
\end{align}
Using eqs.~\eqref{eq:propg1} and \eqref{eq:propg2} we have:
\begin{align}
\int_{-\infty}^{+\infty} x^2 \ g(x) \diff x & = \frac{1}{\lambda^2 (1+\theta) } + \frac{1}{\lambda^2 }.
\end{align}
Finally the variance of $g(x)$ becomes: 
\begin{align}
\sigma_g^2 & = \int_{-\infty}^{+\infty} x^2 \ g(x) \diff x - Q_g^2 \nonumber \\
& = \frac{1}{\lambda^2 (1+\theta) }. \label{eq:sg5}
\end{align}
Now, $Q_f$ and $\sigma_f$ can be computed with the substitutions:
\begin{align}
\lambda & \rightarrow \alpha \nonumber \\
\theta & \rightarrow 0, \nonumber 
\end{align}
and the results are:
\begin{align}
Q_f & = \frac{1}{\alpha } \\
\sigma_f & = \frac{1}{\alpha }.
\end{align}

Armed with the above equations, the mean value of $S(x)$ can be worked out:
\begin{align}
G & = \int_{-\infty}^{+\infty} x \ S(x) \diff x \nonumber \\
& = w \int_{-\infty}^{+\infty} x \ f(x) \diff x + (1-w) \int_{-\infty}^{+\infty} x \ g(x)  \diff x \nonumber \\
& = \frac{w}{\alpha} + \frac{1-w}{\lambda}. \label{gain}
\end{align}
This is, of course, the weighted average between the means of the exponential and gamma distributions. 
Eq.~\eqref{gain} is very important and it can be utilized to measure the gain when the parameters of $S(x)$ are known. 
We will exploit this formula in later sections of this technical note and especially in section~\ref{sec:fit} where the \emph{fitting method} is introduced. 

On the other hand, $\sigma_G$ is difficult to derive and its proof will not be included here.  
Fortunately, a similar calculation has been performed in Ref.~\cite{me3} and it is explained there in great detail.  
Employing eq.~(3.3) of Ref.~\cite{me3} the variance of $S(x)$ can be written as:
\begin{align}
\sigma_G^2 = \frac{w}{a^2} + (1-w)\frac{1}{\lambda^2(1+\theta)} + w(1-w)\left( \frac{1}{\lambda} -\frac{1}{a} \right)^2. \label{rms}
\end{align}
We will not use eq.~\eqref{rms} here, but we decided to include the formula (complicated as it is) mainly for consistency reasons. 
Now we have all the machinery needed to describe the standard techniques for gain calibration. 
We will start with the simplest one: the \emph{occupancy} method.

%
%

\section{Model independent approach}
\label{sec:occ}

\subsection{Basic formulae}

One can solve eq.~\eqref{eq:G} for $G$ and the result is:
\begin{align}
G = \frac{Q_R - Q_0}{\mu}.
\label{eq:occG}
\end{align}
It is obvious from eq.~\eqref{eq:occG} that if one could obtain good estimates for $Q_R$, $Q_0$ and $\mu$ then one can calculate $G$ in the simplest of ways. 
This is the easiest approach to gain determination.  
Now, one can take data with a PMT and a poissonian light source and compute $Q_R$ by the mean of this sample, $Q_R^\prime$. 
In general in this section, and later, we will denote primed parameters as estimates to the true parameters that are usually unprimed. 
The mean of the pedestal, $Q_0$, can be evaluated by taking data with the light source off. 
This is what Saldanha \emph{et al.} call {blank data}~\cite{salda}. 
This distribution, ignoring dark current, is usually a simple gaussian that one can fit and extract both $Q_0^\prime$ and $\sigma_0^\prime$.
Now, the calculation of $\mu$ is the most tricky part of this analysis. 

From eq.~\eqref{eq:pois3} one finds that the probability for a light source to produce no PEs is different from zero and, in fact, equals to:
\begin{align}
P(0;\mu) = \mathrm{e}^{-\mu}. 
\label{eq:prob3}
\end{align}
One can solve eq.~\eqref{eq:prob3} for $\mu$ and the result is:
\begin{align}
\mu = - \mathrm{log} ( P(0;\mu) ).
\label{eq:mu1}
\end{align}
It is obvious, from our previous discussion, that $P(0;\mu)$ is given by the ratio of the integral of the pedestal peak ($I_0$), divided by the integral of $S_R(x)$ ($I$). 
\begin{align}
P(0;\mu) = \frac{I_0}{I}
\end{align}
Again, $I$ can be calculated from the SPE spectrum with light on.
The difficult part is to obtain a good estimate for $I_0$, but assuming that (somehow) one derives $I_0^\prime$, the  formula for $\mu^\prime$  becomes:
\begin{align}
\mu^\prime = - \mathrm{log} \left( \frac{I_0^\prime}{I^\prime} \right).
\label{eq:mu2}
\end{align}
One can then get a primed version of the original eq.~\ref{eq:occG}. 
\begin{align}
G^\prime = \frac{Q_R^\prime - Q_0^\prime}{\mu^\prime}
\label{eq:occG2}
\end{align} 
This formula can be used to estimate the gain $G$.

The ratio of $I_0/I$ is commonly known as the pedestal occupancy, or simply \emph{the occupancy}.\footnote{%
Note though that there is no consensus among physicists on how to define occupancy. 
For instance, Becker-Szendy \emph{et al.} define the occupancy as the probability to detect one PEs or more~\cite{IMBcalib}, 
while Saldanha \emph{et al.} identify the occupancy with the poissonian mean $\mu$~\cite{salda}.}
Likewise, the technique that we have presented (for gain determination) will be termed as the occupancy method. 
It was known to physicists for a long time. 
For instance, Becker-Szendy \emph{et al.} employed a variant of the occupancy method using pulsed data with increasing light intensity~\cite{IMBcalib}. 
This technique has been recently revived by R. Saldanha \emph{et al.} in Ref.~\cite{salda}. 
In particular, Saldanha \emph{et al.} put the method on a sound footing, treating statistical uncertainties and justifying the need for another publication. 
They also offered a novel way to estimate $I_0$ that we now describe. 

\begin{figure}[!t]
\centering
\includegraphics[width=7.5cm, height=5.25cm]{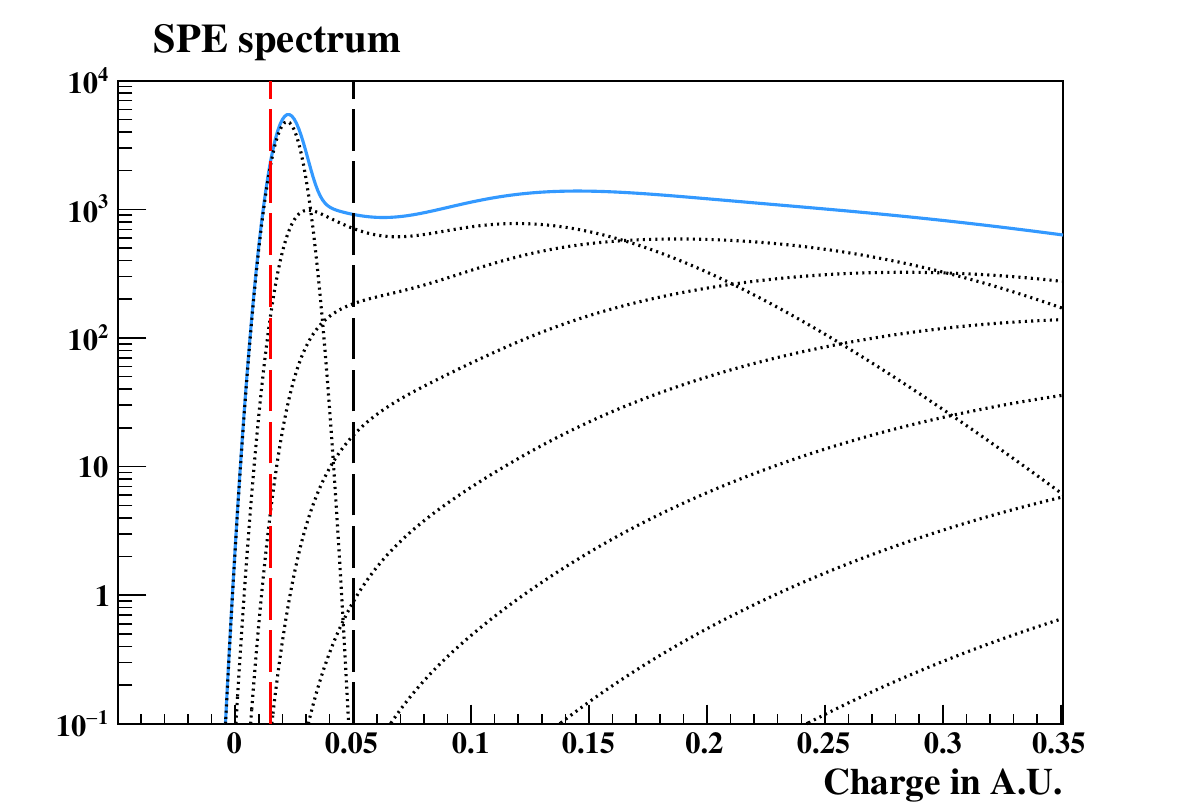} \\[1.5ex]
\includegraphics[width=7.5cm, height=5.25cm]{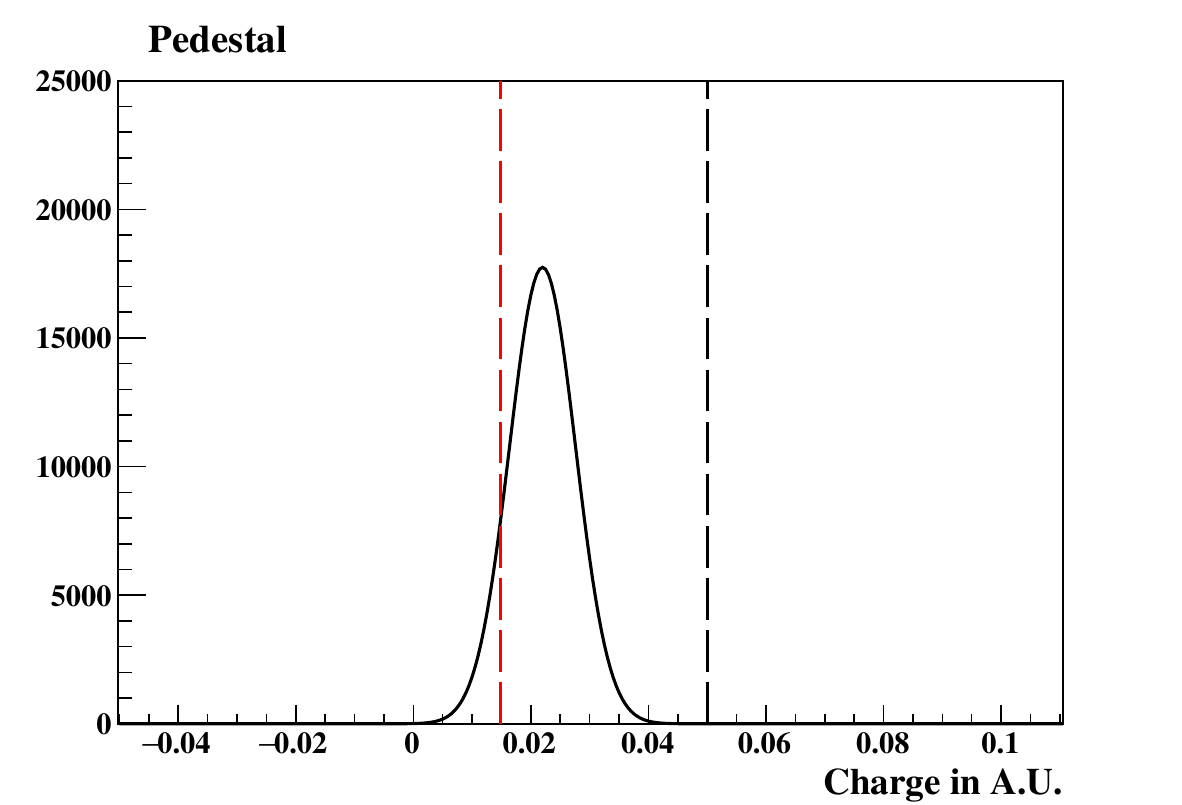} \\[1.5ex] 
\includegraphics[width=7.5cm, height=5.25cm]{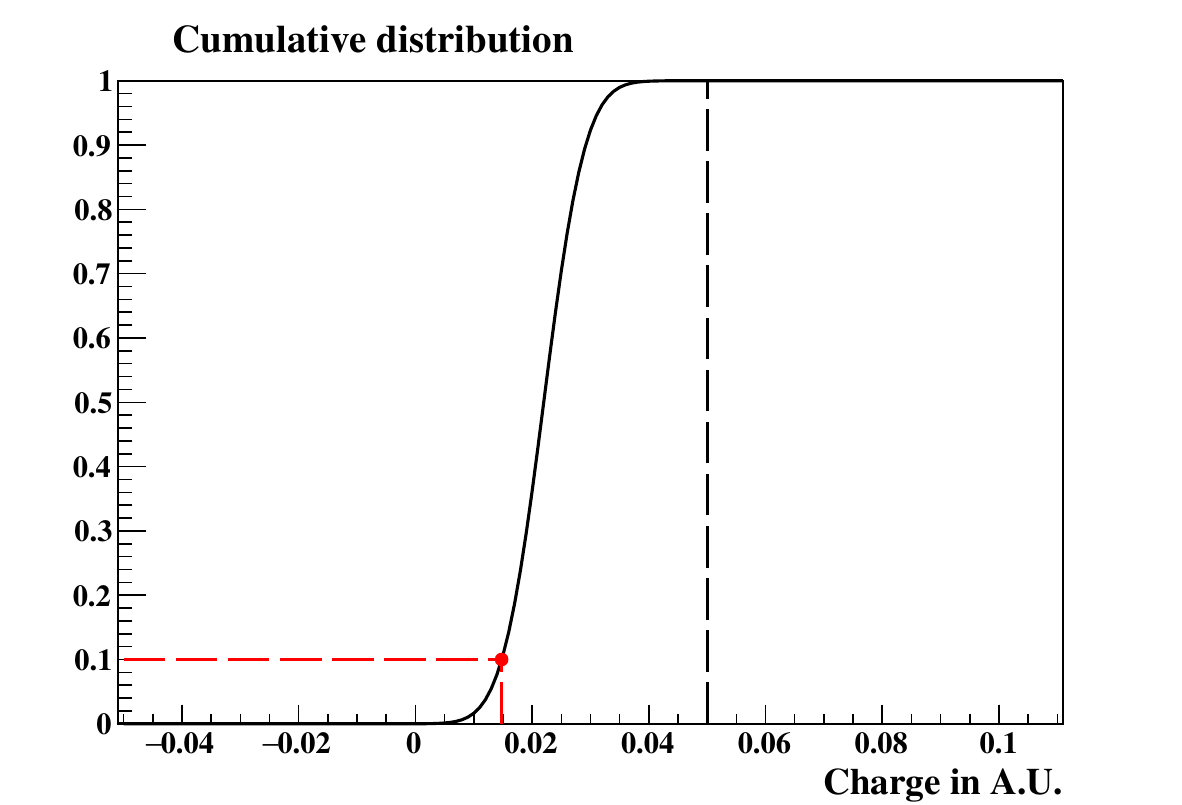} 
\caption{Fig.~(top) shows a zoomed plot of the SPE distribution. The black dashed line shows the $x=Q_0^\prime + 5\sigma_0^\prime$ limit, 
while the red line shows the $x$ value that includes 10~\% of the pedestal. 
Fig.~(middle) shows the pedestal, with the same black and red lines, and  Fig.~(bottom) the cumulative distribution of the pedestal.  }
\label{fig:cuts}
\end{figure}

The simplest way to evaluate $I_0^\prime$ would be to integrate the SPE distribution from $-\infty$ to $x=Q_0^\prime + 5\sigma_0^\prime$.\footnote{%
Actually, one can start the integration from $x=Q_0^\prime - 5\sigma_0^\prime$, since for values of $x<Q_0^\prime - 5\sigma_0^\prime$ the gaussian distribution falls to zero. }
This limit is shown in the vertical, dashed black line of the zoomed SPE spectrum, Fig.~\ref{fig:cuts} (top). 
As it is shown in this figure, all the pedestal is included in this range (first dotted line). 
Nonetheless there is a serious drawback. 
In particular, the contributions of at least three of the first PEs fall in this window as well, as it is shown in Fig.~\ref{fig:cuts} (top). 
Integrating from $-\infty$ to $Q_0^\prime + 5\sigma_0^\prime$ will undoubtedly overestimate $I_0$. 
To bypass this obstacle Saldanha \emph{et al.} proposed to limit the integration to a lower upper limit that includes only a fraction $f$ of the pedestal. 
Say to an $x$ value that includes 10~\%. 
This $x$ value can be calculated with blank data, and it is shown in Fig.~\ref{fig:cuts} (middle) with the red vertical dashed line. 
The cumulative distribution of the pedestal is also shown in Fig.~\ref{fig:cuts} (bottom), and one sees that the red dashed line indeed cuts  at 10~\%. 
Integrating the SPE spectrum from $-\infty$ to the red line, one obtains the truncated integral of $I_T^\prime$. 
This is $f$ times $I_0^\prime$, and the formula holds:
\begin{align}
I_0^\prime = \frac{I_T^\prime}{f}. 
\end{align}
Putting all this together one gets the equation for $\mu^\prime$:
\begin{align}
\mu^\prime = - \mathrm{log} \left( \frac{I_T^\prime}{f I^\prime} \right).
\label{eq:mu3}
\end{align}
This is (in our view) the most important result of Saldanha \emph{et al.}~\cite{salda}. 
Using eqs.~\eqref{eq:occG2} and \eqref{eq:mu3} one gets the gain $G$ in an almost unbiased way.  

\subsection{Generated data}

In this subsection we put the occupancy method to the test. 
In particular, we generated and analyzed a series of toy Monte Carlo (MC) data.  
In this way, the parameters of $B(x)$ and $S(x)$ were known and one knows what to expect. 
The algorithm for generating pseudo experiments has been presented in Ref.~\cite{me3} and it will be repeated here. 
In general, it follows the main steps of PMT operation. 
\begin{enumerate}[i.]
\item First, a number of PEs was thrown from a Poisson distribution of mean value $\mu$.
\item For each PE, a charge was picked randomly from the SPE distribution of eq.~\eqref{eq:polya}, and 
the total charge was calculated by summing the individual charges that each PE deposits.
\item A charge was thrown from the gaussian distribution of the pedestal and summed to the total charge obtained in step two.
\item The total charge was filled in a histogram.
\item This procedure was repeated for 2.5 $10^5$ times and a histogram of 2.5 $10^5$ entries was produced.
\end{enumerate}
For all the MC studies included in this document, a range of a thousand bins between -0.6395  and 1.3605  was assumed. 
These values give a bin width of 0.002.  
This peculiar binning scheme was chosen to match the format of the data analyzed in Ref.~\cite{me2}. 

\begin{figure}[!t]
\centering
\includegraphics[width=7.5cm, height=5.25cm]{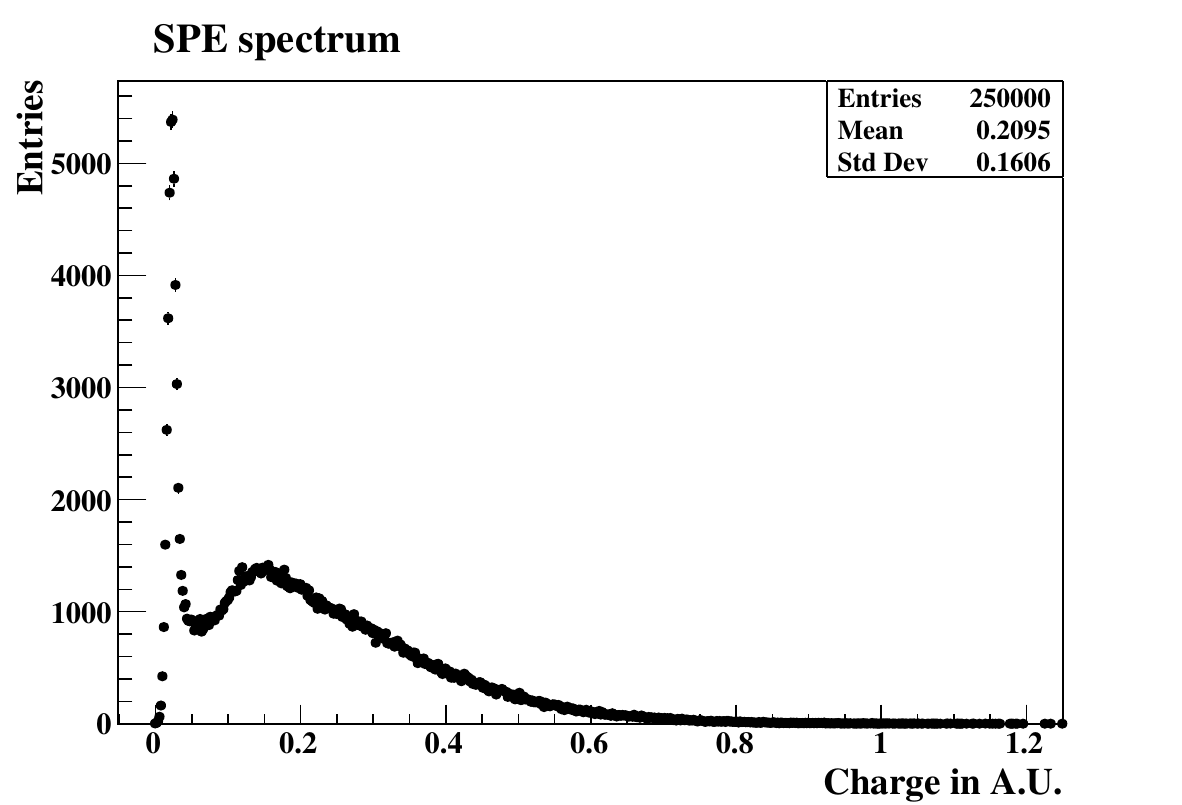} 
\caption{A toy Monte Carlo SPE histogram.}
\label{fig:toy}
\end{figure}

Fig.~\ref{fig:toy} shows the SPE spectrum of a single pseudo experiment. 
This histogram was generated with $\mu=2$ and the parameters of $B(x)$ and $S(x)$ quoted in section~\ref{sec:spe}. 
Of course, it is directly comparable to the plot of Fig.~\ref{fig:plot}, albeit with the addition of statistical fluctuations. 
In general, here, toy experiments were generated with three different values of $\mu= 1, 2$ and $3$ and for a wide span of threshold fractions between $0.01$ and $0.85$. 
It should be understood, that by lowering the threshold one gets a purer sample of pedestal events (by the integration of the lower tail of the SPE spectrum), but at the same time one gets a smaller sample with larger statistical errors. 
So, there is a synergy in action !
 As a matter of fact, Saldanha \emph{et al.} have proposed to use $f=0.1$ for a wide range of poissonian mean values. 
In this subsection we tried to reproduce this result using a different approach. 
In this process we probed the accuracy of the occupancy method as well. 

For each pair of $(\mu$, $f)$  a series of one thousand MC data were generated. 
It should be noted that, before each toy, a pedestal run was simulated with $10^4$ events.
This histogram was fitted with a simple gaussian to extract $Q_0^\prime$ and $\sigma_0^\prime$.
Knowing the pedestal values one can compute $I_T^\prime$ easily. 
As a matter of fact, $I^\prime_T$ equals to:
\begin{align}
I^\prime_T = N_T \Delta x, 
\end{align}
where $N_T$ is the number of events below the threshold, and $\Delta x$ the bin width of the SPE histogram. 
A similar relation holds for $I^\prime$. 
Forming the ratio $I^\prime_T/I^\prime$ $\Delta x$ drops out, and one is left with the equation:
\begin{align}
\mu^\prime = - \mathrm{log} \left( \frac{N_T}{f N_{tot}} \right),
\label{eq:mu4}
\end{align}
where $N_{tot}$ is the total number of entries of the SPE histogram.
This is the fundamental formula that gives $\mu$. 

Now, armed with $Q_0^\prime$ and $\sigma_0^\prime$ (taken from a pedestal run) one can calculate the $x$ value that cuts at a threshold fraction $f$. 
One can then find $N_T$ and $N_{tot}$ from the statistics of the sample. 
Of course, eq.~\eqref{eq:mu4} will give $\mu^\prime$ and, finally, eq.~\eqref{eq:occG2} provides an estimate for the gain.  
It should be noted, that the occupancy method has been implemented in the \texttt{git-hub} repository \cite{git} using a \texttt{C++} software.
The code is simple and it can provide $G$ with two histograms as input. 
One for blank data, and one for the SPE spectrum with light on. 
The code is fast and flexible in the sense that it can provide the gain $G$ for any given input value of $f$. 
It usually takes less than a second to analyze a single case.

\begin{figure}[!t]
\centering
\includegraphics[width=7.5cm, height=5.25cm]{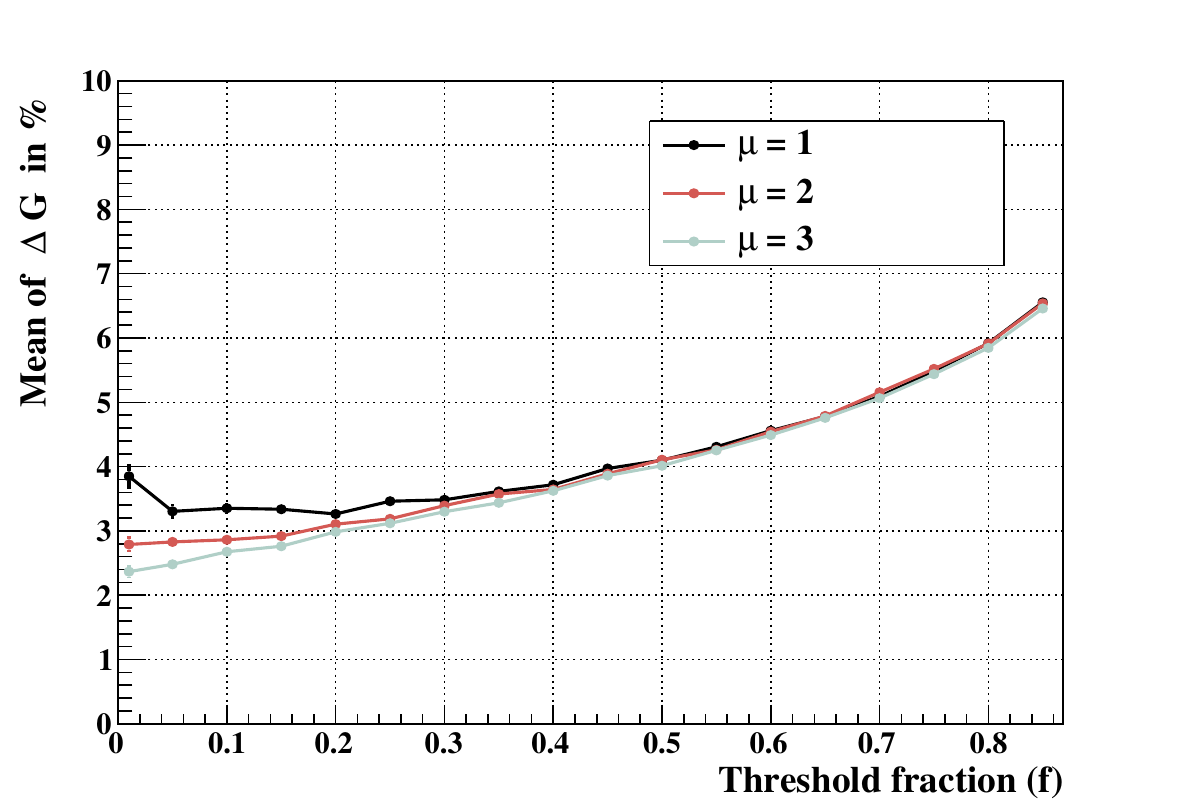} \\[1.5ex]
\includegraphics[width=7.5cm, height=5.25cm]{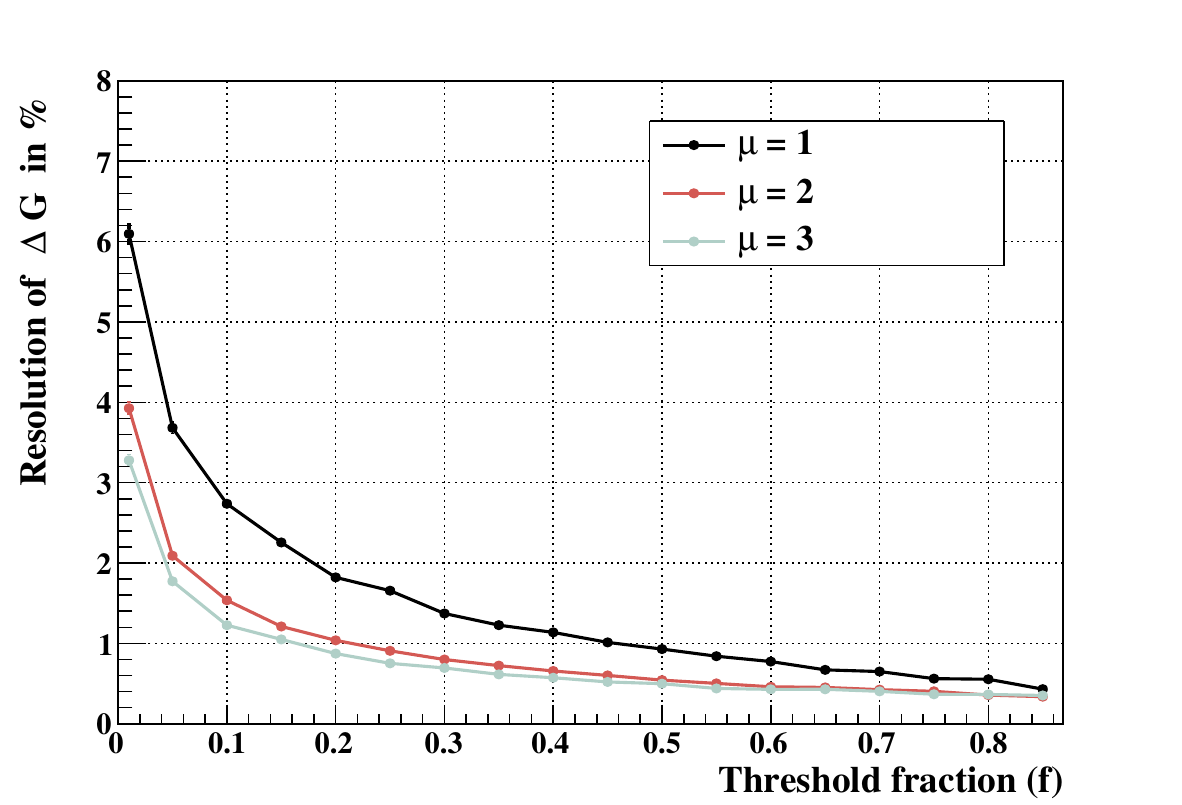} 
\caption{Mean value of the fractional deviation from the true gain (top) and its RMS (bottom).}
\label{fig:ores}
\end{figure}

Using this software we analyzed an extensive series of MC data. 
As mentioned before, for a particular pair of $(\mu$, $f)$ one thousand pseudo experiments were generated. 
For each toy, the value of $\Delta G$ was computed: 
\begin{align}
\Delta G = \frac{G^\prime - G}{G}.
\label{eq:frac}
\end{align}
$G^\prime$ is the estimate of the gain (taken from the occupancy technique), and $G$ is the true value of the injected gain. 
$\Delta G$ is of course the fractional deviation from the true gain. 
As calculated in the previous section $G$ equals to:
\begin{align}
G = \frac{w}{\alpha} + \frac{1-w}{\lambda},
\end{align}
giving a true value of $G=0.0936$ using the numbers of section~\ref{sec:spe}. This was the same for all pseudos. 
Now for every simulated experiment, $\Delta G$ was filled in a histogram. 
This was repeated giving at the end of the run a plot of $10^3$ entries. 
Fig.~\ref{fig:ores} (top) shows the mean of this distribution for different values of $\mu$ and $f$, while Fig.~\ref{fig:ores} (bottom) shows the RMS.
A few remarks are necessary here. 

First, as $f$ increases so does the fractional deviation. Of course, we expected this behavior from our previous discussion.  
Nonetheless, it is interesting that for $f>0.4$ one finds the same difference for all values of $\mu = 1, \ 2$ and $3$, Fig.~\ref{fig:ores} (top). 
Around $f = 0.1 - 0.2$ the curves of the mean have a minimum. 
For $f=0.2$, $\Delta G$ of all three curves is around 3~\% with an RMS between 1~\% and 2~\%. 
In this note, we will analyze MC and detector data using a threshold fraction of  0.2 since it gives smaller RMS, while the fractional deviation is not far from the $f=0.1$ case.
In any case, this choice is not very different from what Saldanha \emph{et al.} suggest~\cite{salda}.

%
%

\section{Conventional method}
\label{sec:fit}

The occupancy method has three main advantages:
\begin{enumerate}[i.]
\item First, it is simple and straightforward. 
\item Second, it is fast. Usually it takes less than a second to analyze a single case, and
\item Third, one does not need to assume a model for the SPE response $S(x)$. 
\end{enumerate}
For all its merits though, it does not come without flaws. 
In particular, we showed in the previous section that it computes the gain with a bias of roughly 3~\%. 
Second, it cannot determine the SPE charge response function that is necessary for several studies as explained in section~\ref{sec:basic}. 

Some of these points can be addressed by the conventional approach to gain calibration. 
This is the original method proposed by Bellamy and collaborators back in 1994~\cite{Bellamy}.
The main idea behind it is rather simple. 
Take pulsed light data with a PMT and record the charge distribution. 
Now, these data should be taken at a light level where only of a small number of PEs are registered; around the so-called \emph{SPE level}.
Ideally, a poissonian mean between (roughly) 0.5 and 3.0 should do the trick.
Then fit this distribution with a model for $S_R(x)$ and extract the values of the pedestal, $\mu$ and all the parameters of the SPE response function $S(x)$.  
To do this, of course, one has to pick a model for $S(x)$. This depends on the PMT under examination. 

For this purpose, a standard gaussian $\chi^2$ (that compares data to the model) was built:
\begin{align}
\chi^2 = \sum_i \frac{ (D_i - S_{R}(x_i) )^2 }{D_i}.
\end{align}
$D_i$ is number of entries in the $i$-th charge bin and $S_{R}(x_i)$ the prediction. 
The minimum of $\chi^2$, the so-called \emph{best fit}, was found using the \texttt{Minuit2} package~\cite{minuit2}. 
Now, this fit is rather complicated and it is likely to fail unless good initial values and limits are input in the minimizer. 
To account for the normalization of the charge spectrum we multiply $S_R(x)$ with a factor $N$ times the bin width. 
Like this, $N$ should be approximately equal to the number of entries of the SPE histogram ($N_{tot}$). 
We insert a value of $N=N_{tot}$ in the minimizer and constrain it between $0.75~N_{tot} - 1.25~N_{tot}$. 

For all the histograms treated in this study, a simple gaussian fit was first performed around the maximum of the distribution to extract the normalization ($N_0^\prime$), 
the mean ($Q_0^\prime$) and standard deviation ($\sigma_0^\prime$) of the pedestal. 
$Q_0$ was set equal to $Q_0^\prime$ and fixed between $Q_0^\prime - 0.5~\sigma_0^\prime$ and $Q_0^\prime + 0.5~\sigma_0^\prime$.
On the other hand, $\sigma_0$ was set to $\sigma_0^\prime$ and limited between $0.5~\sigma_0^\prime $ and $1.5~\sigma_0^\prime$.
An estimate of the mean number of PEs was given by the formula: 
\begin{align}
\mu^\prime = - \log \left(  \frac{N_0^\prime}{\ N_{tot}} \right),  
\label{eq:muestim}
\end{align}
and was input as an initial value to aid the minimizer. 
Eq.~\eqref{eq:muestim} should be obvious for our previous discussion on the occupancy method.
Note, that we could have computed $\mu$ as in the occupancy, using a threshold, but this was not necessary in our work. 
Note also, that in this way there is no need for blank data.
Last, $\mu$ was constrained between $0.5~\mu^\prime$ and $2.0~\mu^\prime$. 

In this section, just like the previous one, we analyzed MC samples to assess the accuracy of the fitting method. 
Data were generated according to the gamma distribution model of eq.~\eqref{eq:polya}. 
Likewise, $S_R(x)$ was computed using the same model for $S(x)$. 
The final function for $S_R(x)$ was derived numerically using the DFT machinery developed in section~\ref{sec:num} 
since the mathematics involved in the calculations of the $S_n(x)$ and $(S_n*B)(x)$ convolutions cannot be solved analytically. 
Of course, one needs good initial values for $\lambda$, $\theta$, $\alpha$ and $w$. 
The initial value of $\lambda$ was set to: 
\begin{align}
\frac{\mu^\prime}{ Q_R^\prime - Q_0^\prime },
\label{eq:laestim}
\end{align}
where $Q_R^\prime$ is the mean of the charge histogram as calculated from the sample. 
To understand this choice, we note that the average value of $S(x)$ (that is the gain, $G$) is equal to: 
\begin{align}
G = \frac{w}{\alpha} + \frac{1-w}{\lambda}. 
\label{eq:G2}
\end{align}
$\alpha^{-1}$ is the mean of the exponential part and $\lambda^{-1}$ the mean of the gamma distribution. 
On the other hand, the average of $S_R(x)$ has been calculated to be:
\begin{align}
Q_R = Q_0 + \mu G.  
\label{eq:x}
\end{align}
Approximating unprimed parameters with primed ones, one gets the formula: 
\begin{align}
G \simeq \frac{Q_R^\prime - Q_0^\prime}{\mu^\prime}. 
\end{align}
We see that putting $\lambda$ equal to $G^{-1}$ one gets the correct order of magnitude.
$\lambda$ was only loosely constrained. 

On the other hand, $\theta$ was initialized to 7.0 and constrained between the large limits of 0.0 -- 56.0. 
The standard deviation of the gamma distribution is: 
\begin{align}
\sigma = \frac{1}{\lambda\sqrt{1+\theta}}. 
\end{align}
Letting $\theta = 7$, it yields a relative $\sigma$ over the $\lambda^{-1}$ mean of $\sim$ 35~\% which is close to what most PMTs have. 
$\alpha$ was set equal to:
\begin{align}
2.0 \ \frac{\mu^\prime}{Q_R^\prime - Q_0^\prime},
\end{align}
since it has to be larger than $\lambda$.  
Last, $w$ was fixed to 0.2 and limited between 0.0 and 0.6 (since for every well-functioning 
PMT the probability of badly amplified PEs cannot be larger than 60\%). 
Now it should be repeated that the aforementioned limits and initial values are absolutely necessary.  
The multidimensional fit of $S_R(x)$ is very complicated and it is likely to fail unless good initial values are inserted in the minimizer. 
Note also that the fit, when successful, subtracts \emph{statistically} the pedestal through the determination of $Q_0$, $\sigma_0$ and allows for a precise evaluation of all the parameters of $S(x)$. 

\begin{figure}[!t]
\centering
\includegraphics[width=7.5cm, height=5.25cm]{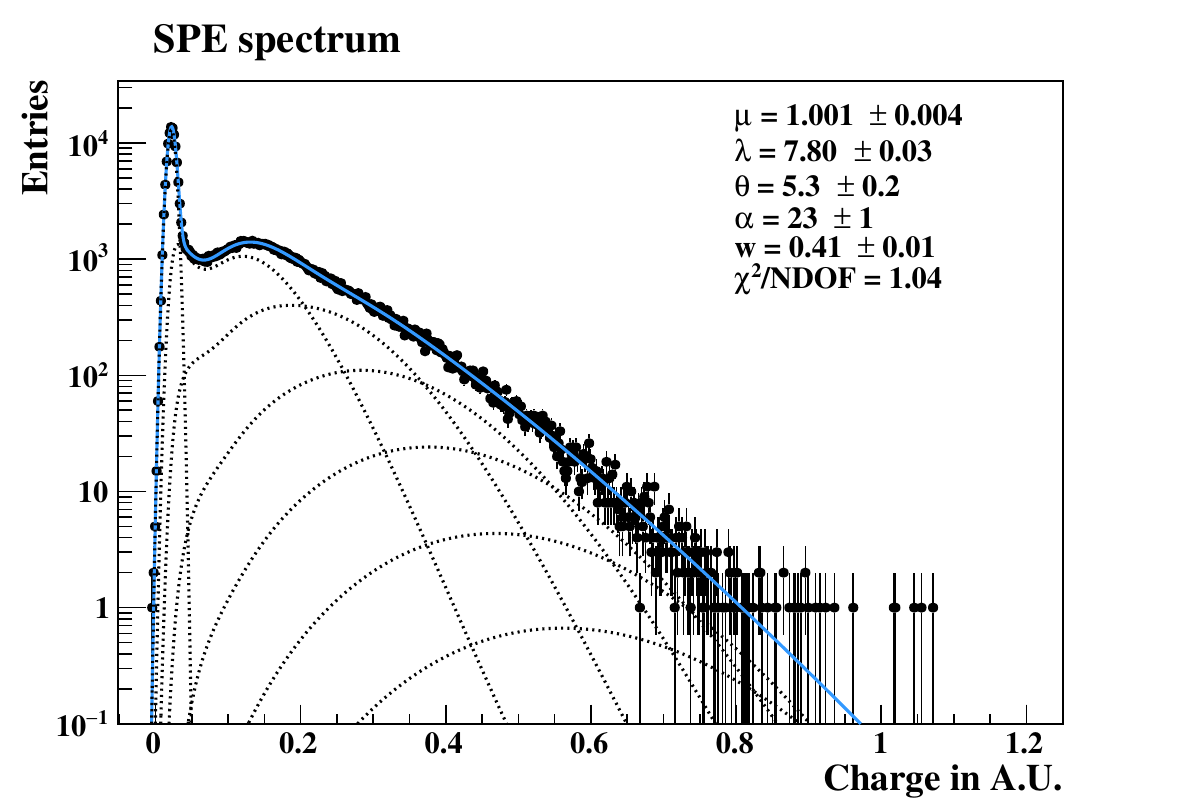} \\[1.5ex]
\includegraphics[width=7.5cm, height=5.25cm]{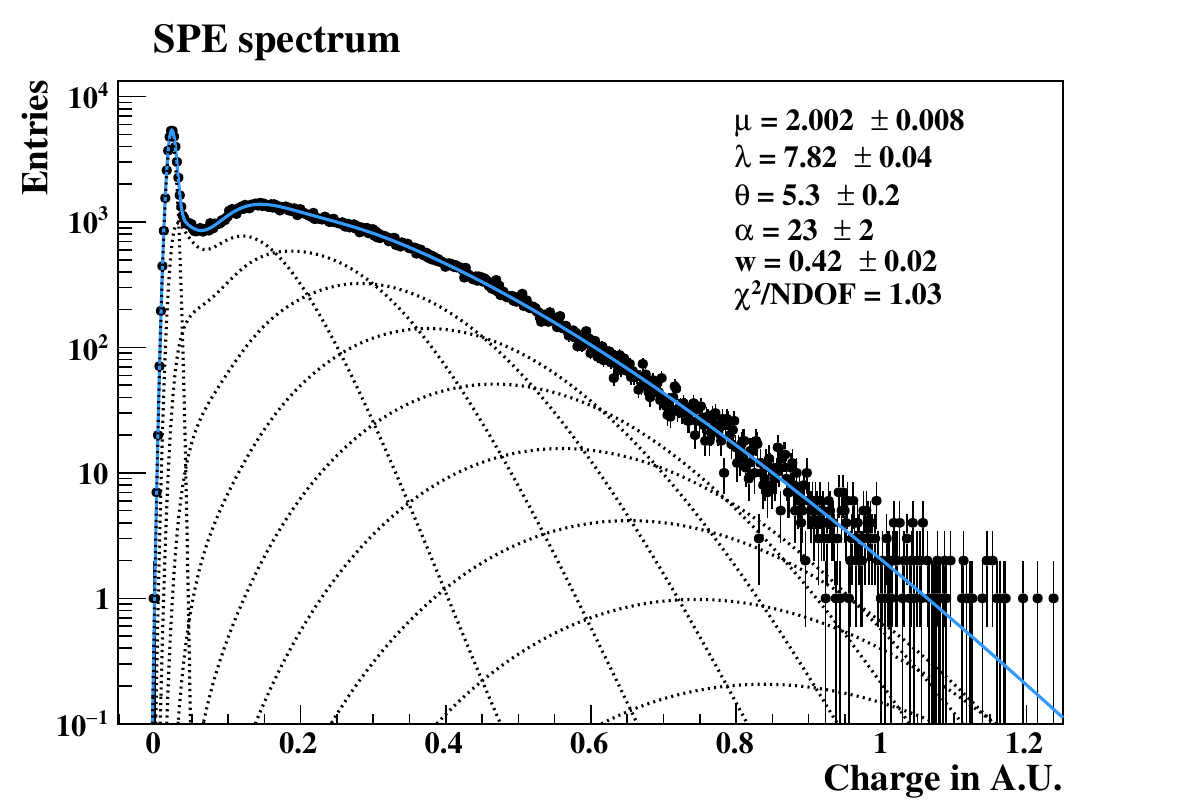} \\[1.5ex]
\includegraphics[width=7.5cm, height=5.25cm]{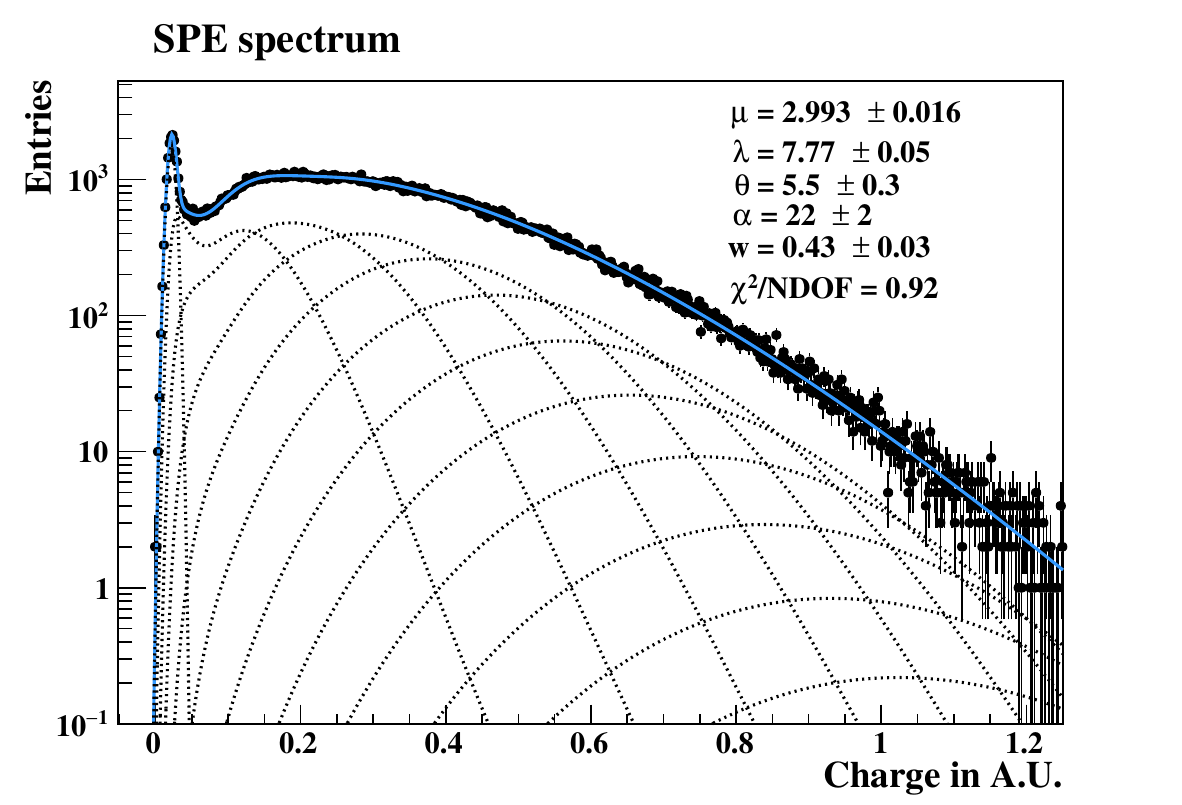} 
\caption{Three fits performed with the $S_R(x)$ function as computed through the DFT technique. 
Black points show the simulated data, and the azure lines the best fit curves. Data were generated for $\mu=1$ (top),  $\mu=2$ (middle) and $\mu=3$ (bottom).}
\label{fig:fits}
\end{figure}

Fig.~\ref{fig:fits} shows three fits done with toy MC data and $\mu=1$, $2$ and $3$.
The azure lines show the best fit curves that follow closely the simulated distributions.
The $\chi^2$ minimum over the number of degrees of freedom (NDOF) was always close to one. 
The parameters of $S(x)$ are consistent with those presented in section~\ref{sec:spe}; always within one or two $\sigma$s. 
Now, data were generated with several values of $\mu$. 
As in previous section 1\,000 toy spectra were produced with $2.5~10^5$ entries each. 
Again the $\Delta G$ distributions were drawn and the mean value of $\Delta G$ was plotted as a function of $\mu$.

\begin{figure}[!t]
\centering
\includegraphics[width=7.5cm, height=5.25cm]{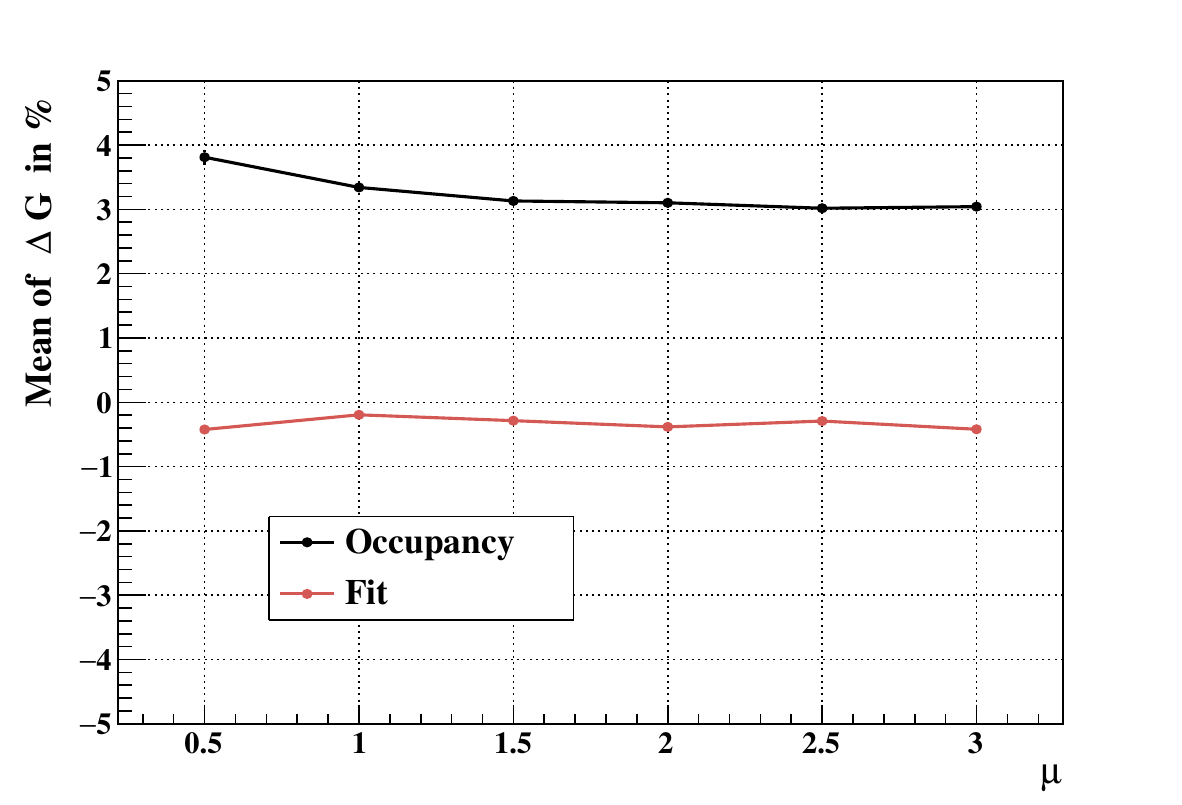} 
\caption{Mean value of the fractional deviation from the true gain, for the occupancy (black) and the fitting method (red).}
\label{fig:fitr}
\end{figure}

Fig.~\ref{fig:fitr} shows the results of this exercise. 
Data were analyzed both with the fitting method and the occupancy. 
Black points show the results of the occupancy and red points those of the fitting technique.  
One sees that the fitting method outperforms the occupancy in a wide range of $\mu$s, having an accuracy of better than $\sim$\,0.5 \%. 
The occupancy has a mean value of $\Delta G$ around $3.0 - 4.0$~\% in accordance with the numbers of previous section.  

An important shortcoming of the original Bellamy \emph{et al.} method is the choice of the SPE response model. 
Note that in this section we generated data using the gamma distribution, and fitted them with the same model. 
It was only obvious that we would retrieve the input numbers of section~\ref{sec:spe}. 
In real life though, there can be PMTs where it is not evident from the onset which PDF should be employed.  
And while it is true, that most PMTs can be treated by the Smirnov parameterization, there can be cases where simple gaussians will not work. 
As a matter of fact, Saldanha \emph{et al.} have claimed that it is difficult to find a representative SPER due to the presence of several contributions from underamplified PEs~\cite{salda}. 
Our own (and subjective) view can be summarized in the following bullet points: 
\begin{enumerate}[i.]
\item First, one can always pick a response model and perform the $S_R(x)$ fit to the SPE spectra. 
The quality of the fits in terms of $\chi^2$/NDOF is indicative whether the original assumption is validated or not. Whether the model reproduces the data or not. 
\item Second, and what is even more important. 
One can take data with increasing light intensity and extract from the fit the parameters of the SPE model. 
If they remain stable (within errors) in a wide range of PEs, say between 0.5 to 3.0, then one can be confident enough
that the parameterization describes the data in a faithful way.
\end{enumerate}
We believe these two points to be powerful agents to the task of gain calibration. 
In the following sections, we will put both to the test, using data from a test bench and toy MC using another model for misamplified PEs.

%
%

\section{Data analysis}
\label{sec:data}

In this section we analyzed data taken with a Hamamatsu R1408 PMT model. 
The R1408 PMTs were originally used in the IMB detector~\cite{IMB} and recently in the Double Chooz experiment~\cite{DC}. 
They are known for not having a sharp peak at the single PE position and their calibration through the original Bellamy \emph{et al.} method can be complicated.\footnote{%
In IMB the R1408 PMTs were calibrated by means of the occupancy method, Ref.~\cite{IMBcalib}. }
Here, we process data taken with an experimental apparatus, using both the DFT and occupancy methods. 

First steps towards this study were given in Ref.~\cite{me1}. 
In this reference the exponential part of $S(x)$ that describes PEs missing the first amplification stage was neglected 
due to the lack of an efficient numerical procedure, and only approximate values for the gains were produced.
This was satisfactory for the needs of the Double Chooz Inner Veto calibration. 
Additionally, in Ref.~\cite{me2} a dataset of pulsed light on data was analyzed using the DFT approach to $S_R(x)$. 
To our knowledge, this is the first time that a R1408 PMT has been treated using the fitting method of Bellamy \emph{et al.}
In this note we attempted a re-analysis of the same large statistics sample using a different binning scheme, and different initial values for the $S(x)$ parameters. 
Of course, the numbers that we get are consistent with those of Ref.~\cite{me2}.

\subsection{Experimental apparatus}

\begin{figure}[!t]
\centering
\includegraphics[width=8cm, height=6.45cm]{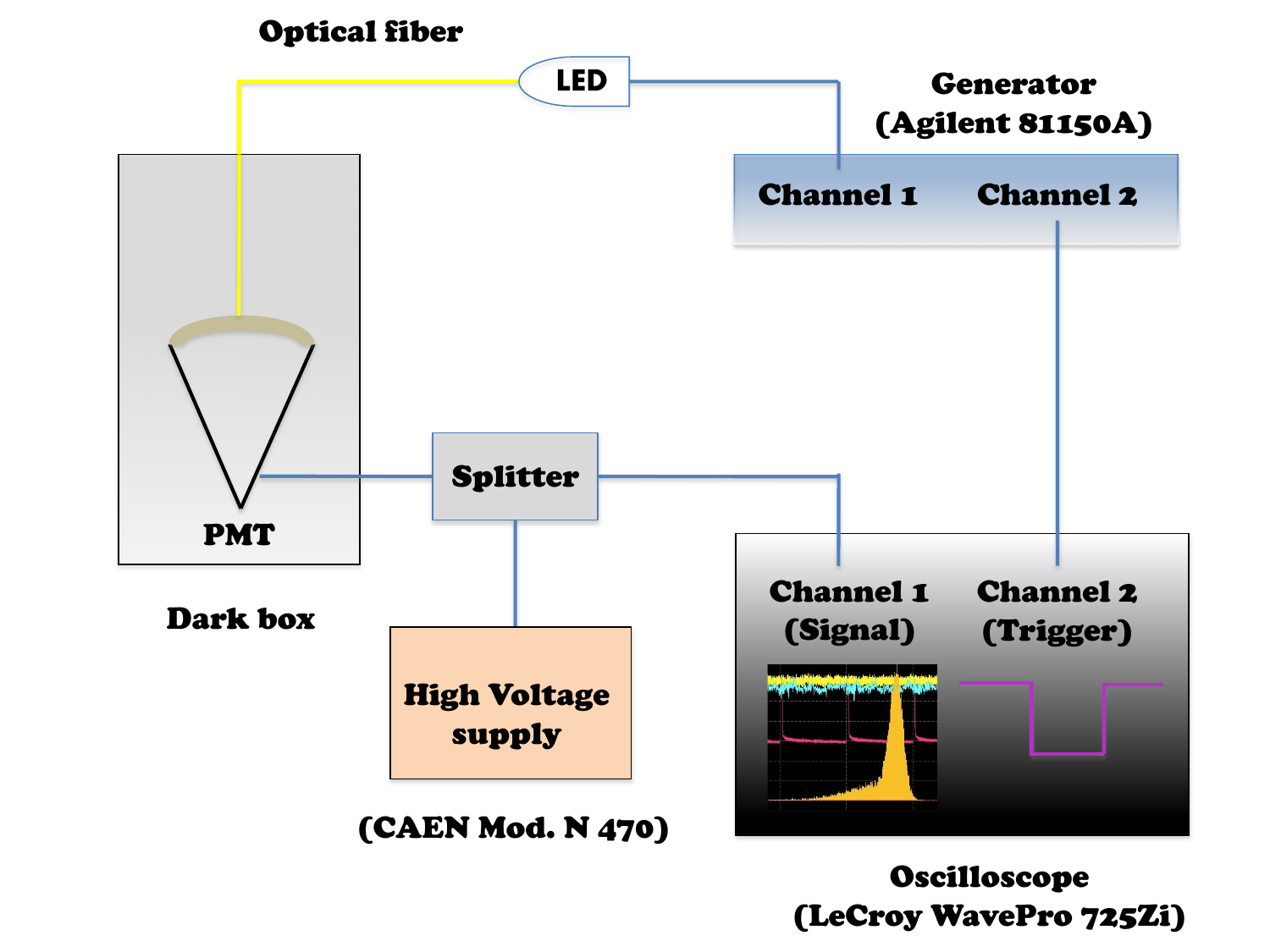} 
\caption{Pictorial representation of the PMT testing apparatus.}
\label{fig:setup}
\end{figure}

The main features of the experimental apparatus needed for the absolute calibration of PMTs are well-known and, more or less, standard.
Most of the components we utilized for this purpose are graphically depicted in Fig.~\ref{fig:setup}. 
As it is shown, the R1408 PMT was placed inside a dark box. 
Extra black silicon was used to ensure that the box was light-tight and the whole setup operated inside a small air-conditioned dark room. 
The temperature of the room was fixed at 21 $^\circ$C. 

The light pulses used to illuminate the PMT were created by a light emitting diode (LED) connected to a fast pulse generator (Agilent 81150A~\cite{agi}). 
The light was directed inside the box and onto the PMT through an optical quartz fiber (Thorlabs BFH48-600~\cite{fib}). 
A halo-like plastic support, attached to the PMT, was used to ensure that the fiber was always on the center of the photocathode and just touching it. 
The same support can be used to change the fiber's position on the photocathode if needed.\footnote{%
More details on the experimental setup, including pictures of the halo-like structure, can be found in ref.~\cite{me1}.}
All of the measurements were done with blue light, $ \lambda = 475$~nm, and the generator operated at 500~Hz. 
The width of the signal pulses were 20~ns.  
The signal from the PMT was separated by the input high voltage through a splitter and was driven to an oscilloscope LeCroy WavePro 725Zi~\cite{lecroy}. 

\begin{figure}[!t]
\centering
\includegraphics[width=7.5cm, height=5.05cm]{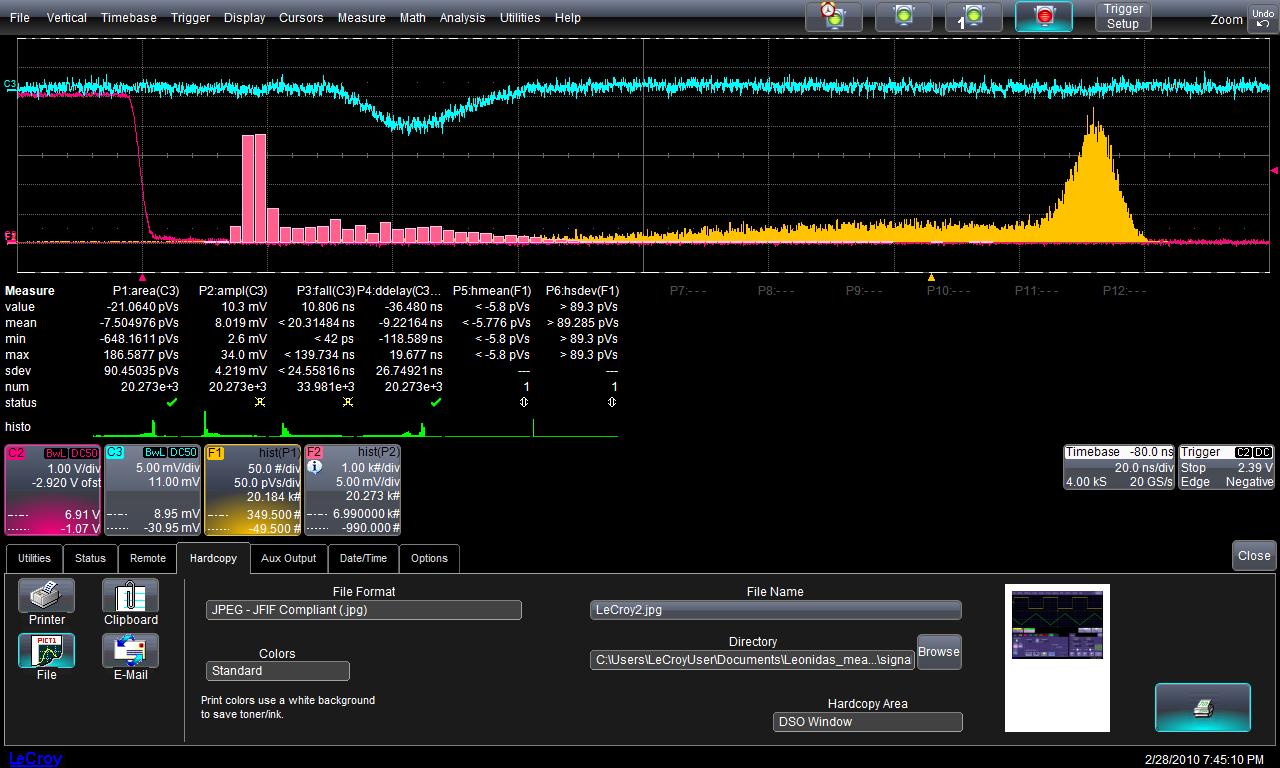} 
\caption{A screenshot from the LeCroy WavePro 725Zi oscilloscope.}
\label{fig:lecroy}
\end{figure}

All datasets were taken with the setup triggering on the generator's second, duplicated, channel sent to the oscilloscope's Channel 2.
This means that no threshold was set in the data acquisition and the charge recorded was integrated in a gate triggered by the generator.  
Fig.~\ref{fig:lecroy} shows a screenshot taken from the oscilloscope. 
The purple curve is the trigger sent by the generator and the cyan curve is the signal of the PMT. 
The two histograms, pink and yellow, refer to the voltage and the charge 
measured on the anode of the PMT respectively. 
In what follows we shall concern ourselves exclusively with the output charge on the oscilloscope; that is the yellow histogram. 
Of course, if no additional noise is present the two histograms are analogous as in Fig.~\ref{fig:lecroy}. 

We should point out that for all the measurements included in this publication the optical fiber was placed at the center of the photocathode 
and that the fiber was always pointing vertically with respect to the surface of the PMT.
In this way one has to deal single quantum and collection efficiencies (that at the center of the PMT) and we expect the assumption of eq.~\eqref{eq:pois3} to be quite valid. 
Note though that the quantum and collection efficiencies of the PMT are expected to vary across the surface of the photocathode and a light source illuminating the whole surface of the PMT will not be 
accurately described by eq.~\eqref{eq:pois3}. In such cases one has to take into account the variance 
across the incident angle and the poissonian factors have to be modified. 
Note also that any possible bias in the extraction of the gain has to be attributed to the validity of the SPE response model (which was taken as an assumption)
and/or the stability of our setup. 

For the purposes of gain determination, the LED input voltage was tuned such that the mean value of observed PEs would lie within the range of $\mu \simeq 0.5$--$3.0$. 
Note that for $\mu \geqslant 0.5$ the contributions from dark current can be neglected as explained in Ref.~\cite{Smirnov}. 
Additionally, for $\mu \leqslant 3.0$ one can clearly observe a pronounced pedestal peak. 
Actually, both methods for gain evaluation (occupancy and fitting) require the presence of the pedestal. 
The occupancy needs the peak to count the number of entries falling below the threshold ($N_T$). 
And, on the other hand, the $S_R(x)$ fitting technique needs a clear pedestal peak since then one can easily deconvolute it.
Moreover, for higher values of $\mu$ spurious correlations might arise that could bias the results.\footnote{%
See for instance the results of table~1 in Ref.~\cite{Bellamy}. }
It has to be noted, that it is always more convenient to limit oneselves in this narrow region since the presence of the pedestal gaussian enforces strong constrains to the fit.
In particular it constraints the poissonian mean $\mu$ through the normalization, $\mathrm{e}^{-\mu}$, of the pedestal. 

\subsection{Results}

To demonstrate that both gain determination methods are self-consistent and that they extract the correct values of $G$ (within their corresponding range of accuracies), 
several datasets were taken with varying light intensity; increasing slowly the LED voltage.
Since the same PMT is used, the two methods (occupancy and fit) should output compatible numbers regardless of $\mu$. 
The pedestal might drift and change, depending on the stability of the setup, but the gain should remain stable. 
This is a powerful condition that can be used to cross-check the validity of results. 

The fitting method offers an additional leverage. 
In particular, it allows for an accurate deconvolution of the SPE charge amplification probability distribution. 
This means that one can calculate detailed values for all the parameters of $S(x)$. 
Of course and (if all goes well) they should remain the same within errors; say within two or three $\sigma$s.

\begin{figure}[!t]
\centering
\includegraphics[width=7.5cm, height=5.25cm]{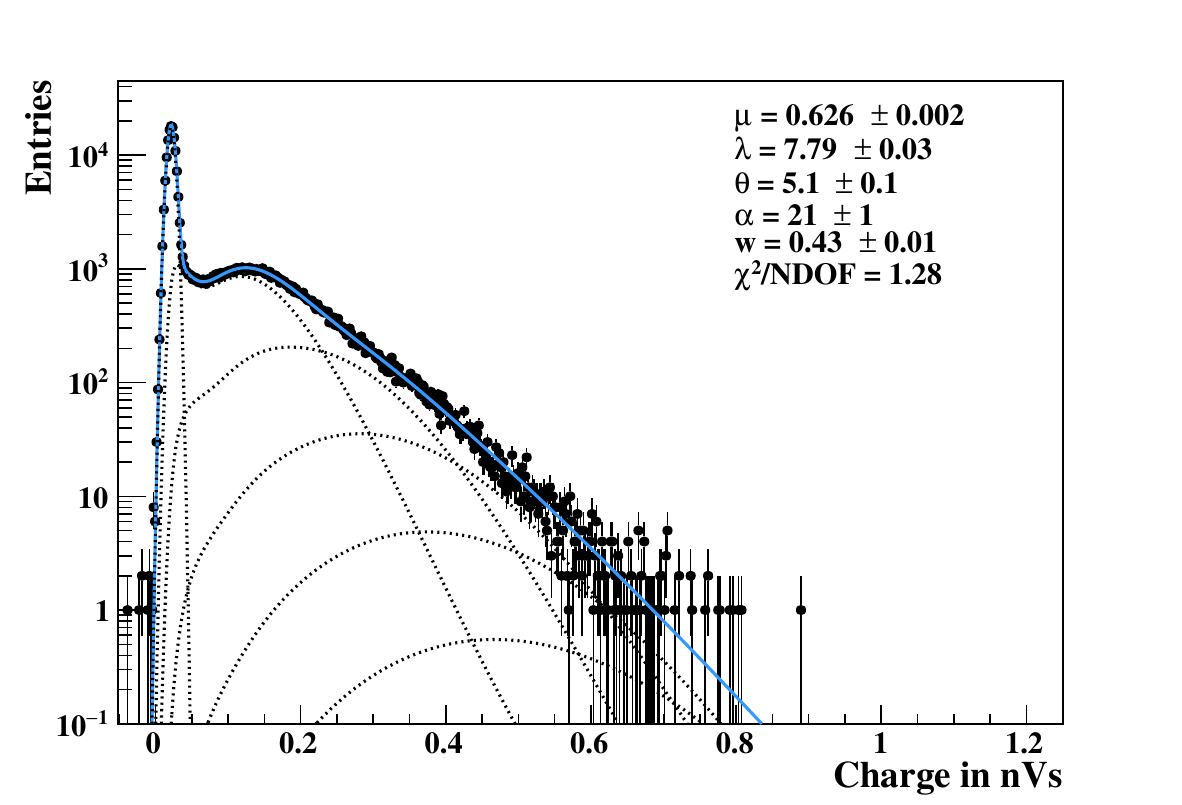} \\[1.5ex]
\includegraphics[width=7.5cm, height=5.25cm]{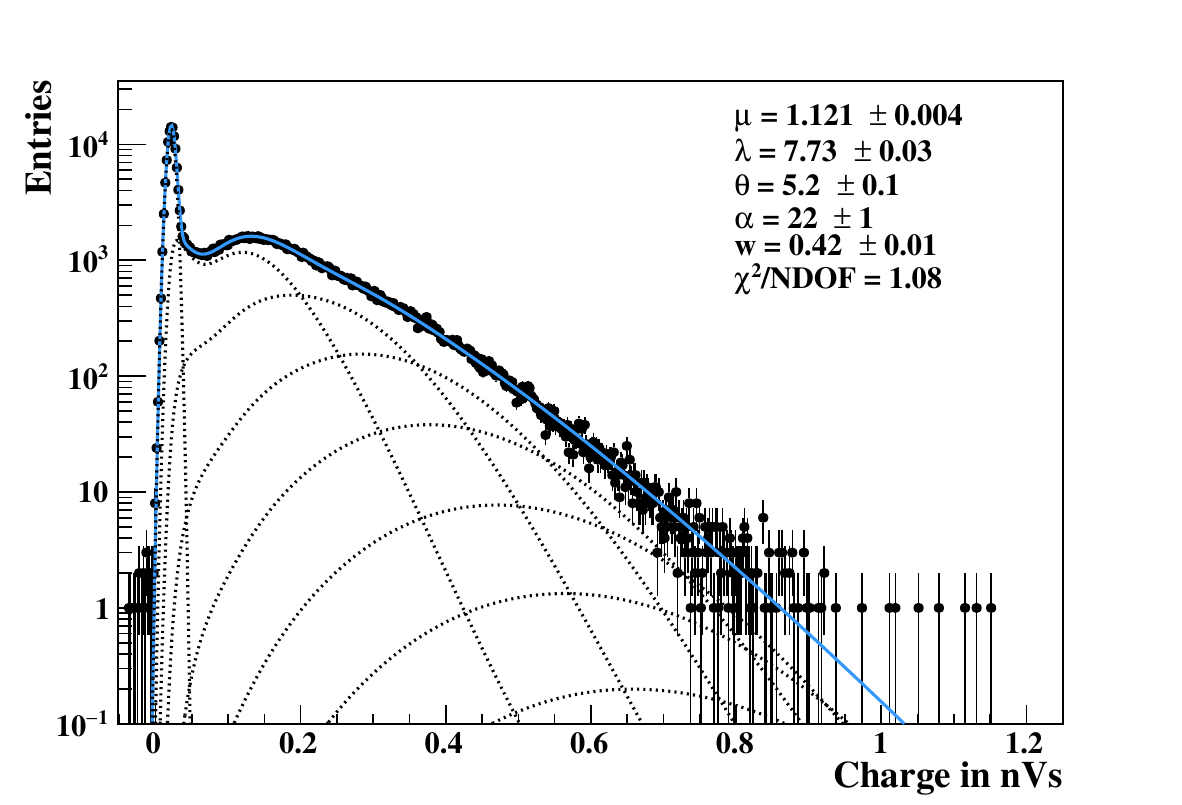} \\[1.5ex]
\includegraphics[width=7.5cm, height=5.25cm]{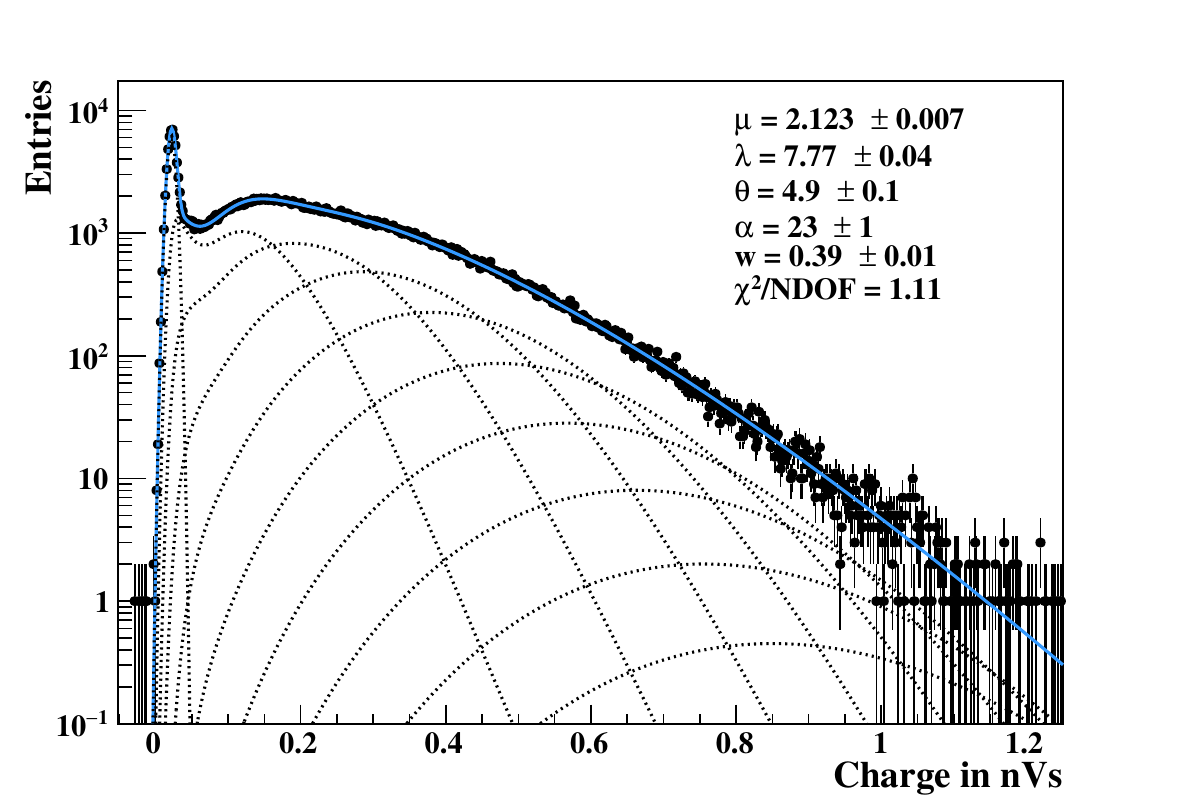} 
\caption{Three fits to data taken with a R1408 PMT. }
\label{fig:fits_data}
\end{figure}

We took fifteen datasets of $\sim$ 2.5 $10^5$ entries each. Before each set a pedestal run of $10^4$ entries was also taken. 
This was used to find estimates for $Q_0$ and $\sigma_0$; as they are needed for both methods. 
The LED voltage was set to values that would correspond to a $\mu$ between roughly 0.5 and 3.0.\footnote{%
This was done by simply counting the zeros.}
Now, it has been said before that the function of section~\ref{sec:spe} can treat the R1408 PMTs~\cite{me1,me2}. 
Likewise, we utilized the gamma distribution model to compute $S_R(x)$. This was done using the DFT numerical approach. 
Fig.~\ref{fig:fits_data} shows three fits taken for different values of $\mu$. 
The data are shown in the black points while the azure lines depict the best-fit curves. 
The quality of the fits was very good as the $\chi^2$/NDOF was always close to one. 
Note also, that the $x$ axis is measured in nVs that is the value of charge given by the LeCroy oscilloscope. 

\begin{table*}[t!]
\centering
\begin{tabular}{| c  || c | c | c | c || c |}
\hline
$\mu$  &  $w$ &  $\alpha$ &  $\lambda$ &  $\theta$ & $\chi^2$/NDOF\\[0.6ex] \hline\hline
0.568 $\pm$ 0.002 & 0.41 $\pm$ 0.01 & 23 $\pm$ 1 & 7.79 $\pm$ 0.04 & 5.1 $\pm$ 0.1 & 1.07 \\
0.626 $\pm$ 0.002 & 0.43 $\pm$ 0.01 & 21 $\pm$ 1 & 7.79 $\pm$ 0.03 & 5.1 $\pm$ 0.1 & 1.28 \\ 
0.692 $\pm$ 0.003 & 0.43 $\pm$ 0.01 & 21 $\pm$ 1 & 7.70 $\pm$ 0.03 & 5.4 $\pm$ 0.2 & 0.95 \\ 
0.722 $\pm$ 0.002 & 0.41 $\pm$ 0.01 & 23 $\pm$ 1 & 7.77 $\pm$ 0.03 & 5.1 $\pm$ 0.1 & 1.21 \\ 
0.819 $\pm$ 0.003 & 0.41 $\pm$ 0.01 & 23 $\pm$ 1 & 7.80 $\pm$ 0.03 & 5.0 $\pm$ 0.1 & 1.18 \\
0.924 $\pm$ 0.003 & 0.42 $\pm$ 0.01 & 22 $\pm$ 1 & 7.75 $\pm$ 0.02 & 5.1 $\pm$ 0.1 & 1.08 \\ 
0.966 $\pm$ 0.003 & 0.42 $\pm$ 0.01 & 22 $\pm$ 1 & 7.75 $\pm$ 0.03 & 5.2 $\pm$ 0.1 & 1.14 \\ 
1.121 $\pm$ 0.004 & 0.42 $\pm$ 0.01 & 22 $\pm$ 1 & 7.73 $\pm$ 0.03 & 5.2 $\pm$ 0.1 & 1.08 \\ 
1.309 $\pm$ 0.004 & 0.43 $\pm$ 0.01 & 21 $\pm$ 1 & 7.75 $\pm$ 0.03 & 5.2 $\pm$ 0.2 & 1.08 \\ 
1.379 $\pm$ 0.004 & 0.42 $\pm$ 0.01 & 22 $\pm$ 1 & 7.74 $\pm$ 0.03 & 5.2 $\pm$ 0.1 & 1.17 \\
1.645 $\pm$ 0.005 & 0.40 $\pm$ 0.01 & 23 $\pm$ 1 & 7.77 $\pm$ 0.03 & 5.0 $\pm$ 0.1 & 1.16 \\
1.920 $\pm$ 0.006 & 0.41 $\pm$ 0.01 & 21 $\pm$ 1 & 7.75 $\pm$ 0.03 & 5.0 $\pm$ 0.2 & 1.00 \\ 
2.123 $\pm$ 0.007 & 0.39 $\pm$ 0.01 & 23 $\pm$ 1 & 7.77 $\pm$ 0.04 & 4.9 $\pm$ 0.1 & 1.11 \\
2.579 $\pm$ 0.013 & 0.40 $\pm$ 0.02 & 23 $\pm$ 2 & 7.75 $\pm$ 0.05 & 5.2 $\pm$ 0.2 & 1.16 \\ 
3.074 $\pm$ 0.016 & 0.46 $\pm$ 0.02 & 20 $\pm$ 1 & 7.58 $\pm$ 0.04 & 5.5 $\pm$ 0.3 & 1.11 
\\[0.6ex] \hline\hline 
\end{tabular}
\caption{Summary of the R1408 PMT calibration results.}
\label{tab:money}
\end{table*}

In this way we analyzed all fifteen datasets. 
The results of this exercise are detailed in table~\ref{tab:money}. 
Indeed, one can see that as $\mu$ becomes larger $w$, $\alpha$, $\lambda$ and $\theta$ remain well within the errors. 
A few further comments should be made.  
First, $\lambda$ can be obtained with much better accuracy than $w$, $\alpha$ and $\theta$. 
This is the case because the contribution of badly amplified PEs is mainly constrained from the valley between the pedestal and the SPE peak. 
It appears that the minimizer has more difficulty to identify $w$ and $\alpha$ in this narrow region. 
Nonetheless, consistent results can be obtained. 
In particular, the $w$ pre-factor (the probability that PEs are not amplified in the full dynode chain) stays constant. 
This should be contrasted with Ref.~\cite{Bellamy} where the exponential part of $S(x)$ is attached to the pedestal and $w$ increases with light intensity. 

\begin{figure}[!t]
\centering
\includegraphics[width=7.5cm, height=5.25cm]{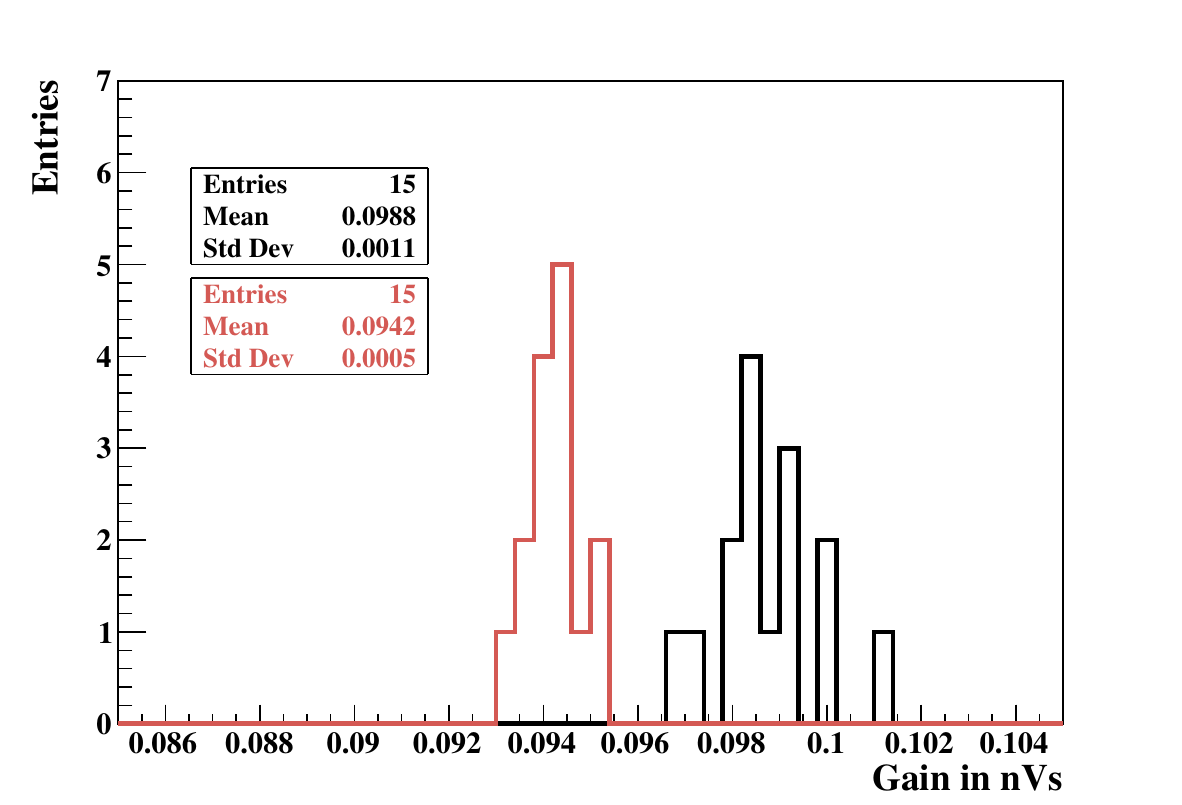} 
\caption{Gain distributions for the  occupancy (black) and the fitting (red) methods.}
\label{fig:Gdist}
\end{figure}

Fig.~\ref{fig:Gdist} shows the one dimensional distributions of the gain for all the measurements of table~\ref{tab:money}. 
The result from the occupancy is shown in the black line and in red this from the fit. 
We repeat that the gain from the fitting  method  was calculated  using  eq.~\eqref{gain}  and  the best-fit parameters for $w$, $\alpha$ and $\lambda$.
On the other hand, that of the occupancy was found employing eq.~\eqref{eq:occG2}. 
Of course, more details about these computations can be found in the appropriate sections. 
Now, the red line is centered around 0.0942 nVs with a standard deviation of better than 1~\%. 
In contrast, the results in black have a larger standard deviation and they differ from the red by $\sim$ 5~\%. 
Note that this is not very far from the value suggested by MC, as shown in section~\ref{sec:fit}.\footnote{%
The fitting and the occupancy method MC results differ by $\sim$ 3.0 - 4.0~\% as shown in section~\ref{sec:fit}.}

The stability of the $S(x)$ parameters versus $\mu$, the quality of the fits in terms of $\chi^2$/NDOF, and the differences with respect to the occupancy method,
point towards the conclusion that the fit can determine the gain in an unbiased way, with an accuracy of better than 1.0~\%. 
We tried to support these assertions with further arguments described in later sections of this document. 

%
%

\section{Soft component}
\label{sec:soft}

\begin{figure}[!t]
\centering
\includegraphics[width=7.5cm, height=5.25cm]{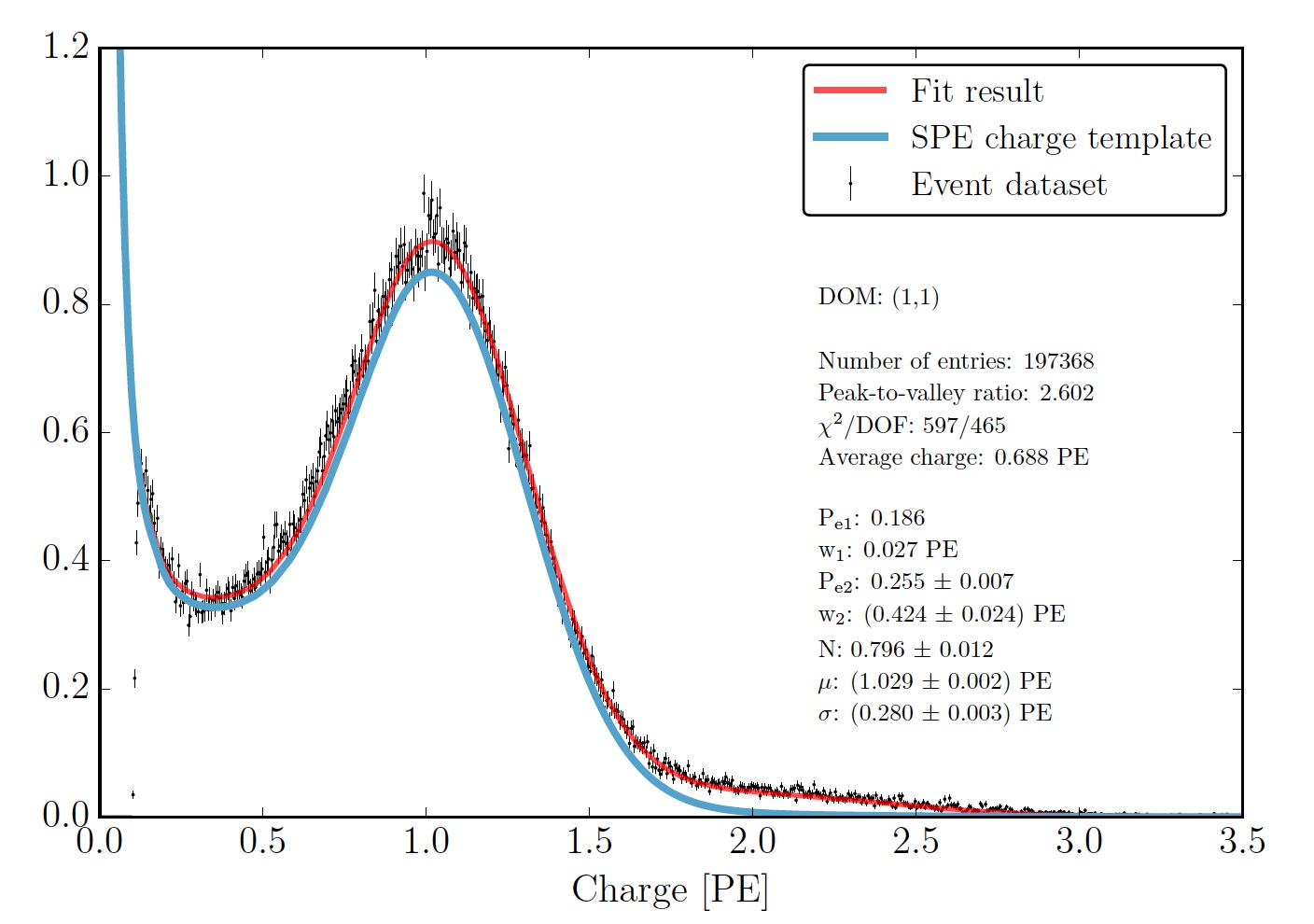} 
\caption{SPE charge response distribution of the IceCube PMTs~\cite{icecube}. }
\label{fig:ic}
\end{figure}

The IceCube collaboration has presented a novel technique to extract the SPE distributions of the PMTs that instrument its detector.  
This analysis is based on a pulse-by-pulse selection of the events that correspond to single PEs, filtering out pedestal and multi-PE signals.  
Note that is an \emph{in situ} measurement, performed while the detector runs. 
More details on this study can be found in Ref.~\cite{icecube}. 
The result for the R7081-02 Hamamatsu model is shown in Fig.~\ref{fig:ic}. 
An interesting byproduct of the IceCube work is the fact that the lower charge part of the SPE response can be best parametrized by a combination of two exponentials. 
The first exponential can be interpreted as PEs missing the first amplification stage. 
On the hand the second, softer component, has no clear origin.\footnote{%
At least no further explanation is given in the IceCube paper~\cite{icecube}.} 
In this section we have tried to investigate whether we can identify this component using the conventional methods for gain determination. 
For this purpose we utilized MC data.   

The gamma distribution model, with the addition of another exponential, becomes:
\begin{align}
S(x) =     &\ w \alpha \mathrm{e}^{-\alpha x }H(x) + w_2 \alpha_2 \mathrm{e}^{-\alpha_2 x }  H(x) \nonumber \\
   & +  ( 1 - w - w_2) \lambda (1+ \theta ) \frac{[ \lambda (1+ \theta ) x ]^\theta }{\Gamma (1+\theta) } \mathrm{e}^{ - \lambda (1+ \theta ) x }  H( x ).
\label{eq:gplus}
\end{align}
Now, the values for $w$, $\alpha$, $\lambda$ and $\theta$ are those presented in section~\ref{sec:spe}. 
On the other hand, the values of $w_2$ and the mean of the second exponential were chosen to be half those of the first exponential distribution $(w,1/\alpha)$. 
In this way we had:
\begin{align}
w_2 = 0.20,
\end{align}
and 
\begin{align}
\alpha_2 = 46.
\end{align}
In our usual way, we generated a series of fake data using the model of eq.~\eqref{eq:gplus} and the values of the parameters explained above. 
Now, the obvious thing would be to fit these simulated histograms with the two exponential gamma model and extract the gain; as it was done in section~\ref{sec:fit}. 
We tried this approach only to fail. 
In particular, this is a fit with ten free parameters and it is very complicated to succeed. 
Most of the fits that we attempted failed or gave a very large $\chi^2/$NDOF. 
It appears that the second exponential is ``hidden'' under the pedestal and the minimizer has no leverage to determine its values. 
To proceed, we went on and analyzed these data using the occupancy method (that requires no model) 
and the fitting technique using the single exponential model of eq.~\eqref{eq:polya}; neglecting thus the second exponential. 

\begin{figure}[!t]
\centering
\includegraphics[width=7.5cm, height=5.25cm]{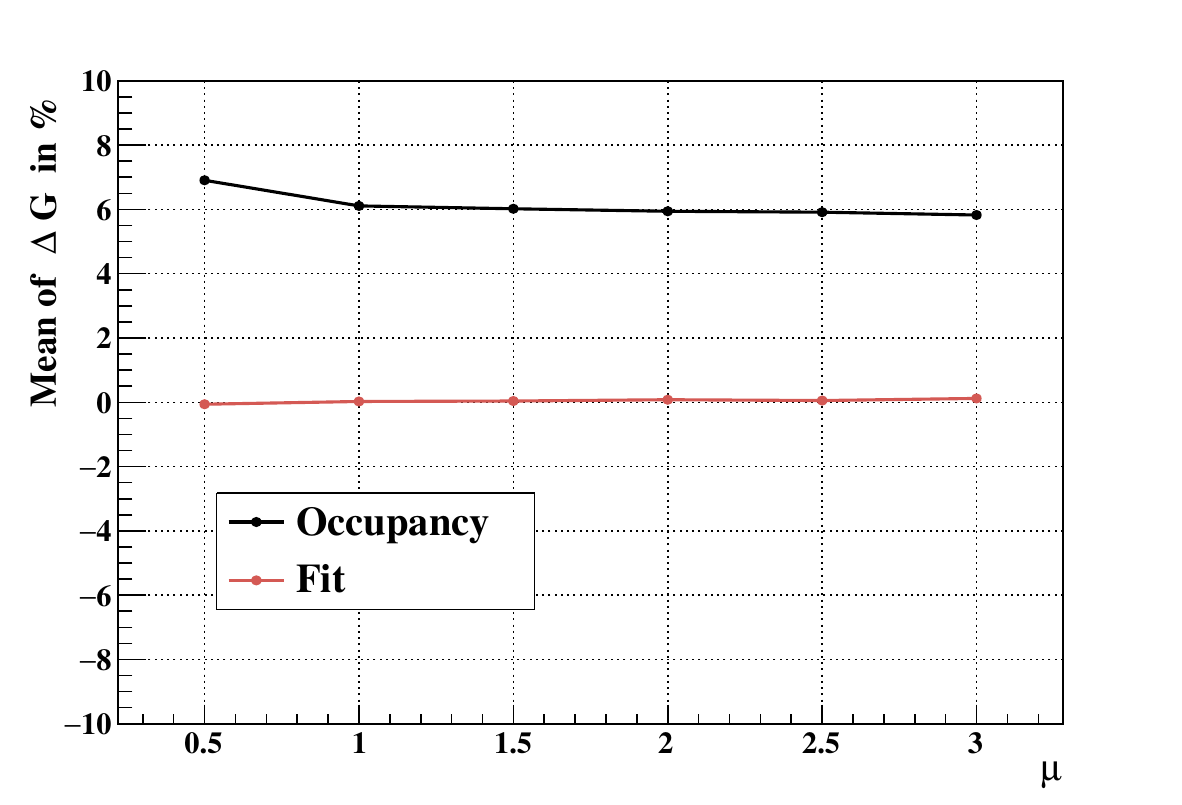} 
\caption{Fractional deviation from the true gain for the occupancy (black) and the fitting method (red). 
These toy MC data were generated according the model of eq.~\eqref{eq:gplus}.}
\label{fig:tune}
\end{figure}

The results of this exercise are shown in Fig.~\ref{fig:tune}.
Now, this plot depicts the mean value of $\Delta G$ versus the true $\mu$. 
$\Delta G$ is of course the fractional difference from the true gain. 
For the fitting method the gain was calculated according to eq.~\eqref{gain}; omitting (again) the second exponential. 
The results are shown with black points for the occupancy and with red those from the fit. 
Several remarks are necessary here. 
First, the occupancy method finds the gain with a bias of $\sim 6.0 \, \%$. 
This is significantly larger than the number found in section~\ref{sec:fit}. 
It appears that the addition of a soft component affects the measurement of $\mu$. 
Note that the occupancy method was proposed as an alternative to the fit, since sometimes it appears to be problematic to choose a representative distribution for badly amplified PEs. 
Nonetheless, we have showed through this simple study that the presence of another exponential degrades the precision of the occupancy. 

Second, and what is even more important. 
The conventional fitting technique calculates $G$ with an accuracy of better than 0.5~\%. 
This is rather surprising ! It appears that the model of eq.~\eqref{eq:polya} can effectively parametrize eq.~\eqref{eq:gplus}. 
In particular, we have seen through individual fits that the values of $w$ and $\alpha$ (those returned from the fit) are larger than the nominal. 
Eq.~\eqref{eq:polya} with these higher numbers provides an excellent approximation of the true SPE distribution. 
Consequently, it is safe to say that the fit is the most precise method for gain determination that we have. 
Of course, these results depend on the values of the parameters inserted in the SPE model. 
Ideally, we would like to try more cases, but this a rather tedious job. 
In any case, we have shown in this simple example that the fitting technique outperforms the occupancy by a large margin. 

%
%

\section{Outlook}
\label{sec:outro}

In this technical note, we presented the main features of the mathematical theory of PMT calibration. 
The standard concepts and principles of PMT operation were described in some detail. 
Furthermore, we worked out the behavior of a PMT in response to a poissonian light source. 
The most common numerical procedures, needed to calculate $S_R(x)$, were laid down, explaining the advantages and disadvantages of each approach. 
Additionally, two methods for gain calibration were put forward. 
One model independent (occupancy) and another one that leans on the choice of a  specific distribution for the SPE response $S(x)$ (fitting). 
We used toy MC data to assess their precision, and we found that the fitting technique is superior  to the occupancy, giving an accuracy of better than 1~\%. 
Real data were analyzed using a R1408 PMT inside a dark and light-tight box. 
Samples were taken with increasing light intensity and we saw that the parameters of the gamma distribution remain stable within experimental errors.  
Finally, we showed that the addition of a soft charge component (attributed to badly amplified PEs) does not affect the results of the fit 
and that a single exponential provides an excellent approximation to the distribution of miss-amplified PEs.

The following remarks are necessary. 
In this note we have only tried to demonstrate the calibration of PMTs in just the simplest case. 
That is, when a light source is placed at the center of the photocathode, and always pointing vertically towards the surface of the PMT. 
These conditions were met both for MC and detector data. 
We have demonstrated that in this simple example, the gain remains remarkably stable (within $\sim 1~\%$ or better) inside the $\mu \simeq 0.5 - 3.0$ plateau. 
We should note that in large (monolithic) detectors equipped with a sizable number of PMTs this simplistic picture ceases to apply. 
In particular, in those circumstances we can expect the extraction of the gain to depend on the geometry of the detector and the position of the event. 
 We can only expect that the accuracy achieved in this article will not be attainable in such cases.
The question of the \emph{in situ} gain calibration of similar, complicated instruments, lies beyond the scope of this publication and was not treated here. 
More details on the energy and spatial resolution of large-volume liquid scintillator detectors can be found elsewhere~\cite{smi1}.


\appendix


\section{Properties of convolution}
\label{app:a}

\subsection{Mean value}

In this appendix we prove two important properties of the convolution. 
First, we show that the mean value of a convolution equals to the sum of the individual means of each distribution. 
Let us take the simple example: $h(x)$ is the convolution of $f(x)$ and $g(x)$.
\begin{align}
h(x) = ( f * g)(x) \label{eq:h}
\end{align}
We demonstrate that the mean value of $h(x)$, E[$h(x)$], is equal to the sum of E[$f(x)$] and E[$g(x)$].
So:
\begin{align}
\textrm{E}[h(x)] & = \int_{-\infty}^{+\infty}  x ( f * g)(x) \diff x \nonumber \\
                         & =  \int_{-\infty}^{+\infty}  x  \int_{-\infty}^{+\infty}  f(y) g(x-y) \diff y \diff x \nonumber \\
                         & = \int_{-\infty}^{+\infty} f(y) \int_{-\infty}^{+\infty} x g(x-y) \diff x \diff y. \label{app:1}
\end{align}
Adding and subtracting the term:
\begin{align}
\int_{-\infty}^{+\infty} y f(y) \int_{-\infty}^{+\infty} g(x-y) \diff x \diff y
\end{align}
eq.~\eqref{app:1} becomes:
\begin{align}
\textrm{E}[h(x)]  = & \int_{-\infty}^{+\infty} f(y) \int_{-\infty}^{+\infty} ( x - y )g(x-y) \diff x \diff y \nonumber \\ & + \int_{-\infty}^{+\infty} y f(y)  \int_{-\infty}^{+\infty} g(x-y) \diff x \diff y. 
\end{align}
Setting
\begin{align}
u = x -y, 
\end{align}
we have:
\begin{align}
\textrm{E}[h(x)]  = & \int_{-\infty}^{+\infty} f(y) \int_{-\infty}^{+\infty} u g(u) \diff u \diff y \nonumber \\ & + \int_{-\infty}^{+\infty} y f(y)  \int_{-\infty}^{+\infty} g(u) \diff u \diff y \nonumber \\
			 = & \int_{-\infty}^{+\infty} f(y) \textrm{E}[g(x)] \diff y +  \int_{-\infty}^{+\infty} y f(y) \diff y \nonumber \\
			 = & \textrm{E}[g(x)] + \textrm{E}[f(x)]. \label{eq:res1}
\end{align}
In these lines we made use of the fact that $f(x)$ and $g(x)$ are normalized to one. 
\begin{align}
\int_{-\infty}^{+\infty} f(y) \diff y = \int_{-\infty}^{+\infty} g(u) \diff u = 1 
\end{align}
The result of eq.~\eqref{eq:res1} can be generalized to the case of a convolution of three, four, etc. distributions. 
For the case of $(S_n*B)(x)$ we have:
\begin{align}
\text{E}[( S_n * B )(x)]  \ = \ & Q_0 + n  G
\end{align}
since there are $n$ convolutions of $S(x)$ and a last one with the pedestal $B(x)$. 

\subsection{Variance}

Second, we prove that the variance of a convolution is equal to the sum of the individual variances of each distribution. 
Let us use the same  $h(x)$ distribution of eq.~\eqref{eq:h}.
We show that:
\begin{align}
\textrm{Var}[h(x)] & = \textrm{Var}[f(x)] + \textrm{Var}[g(x)].
\end{align}
To calculate the variance of $h(x)$ one needs (first) to find the integral:
\begin{align}
\int_{-\infty}^{+\infty}  x^2 h(x) \diff x & = \int_{-\infty}^{+\infty}  x^2 ( f * g)(x) \diff x \nonumber \\
					               & = \int_{-\infty}^{+\infty}  x^2 \int_{-\infty}^{+\infty}  f(y) g(x-y) \diff y \diff x \nonumber \\
					               & = \int_{-\infty}^{+\infty}  f(y) \int_{-\infty}^{+\infty}  x^2 g(x-y) \diff x \diff y.
\end{align}
Then we add and subtract the terms:
\begin{align}
& \int_{-\infty}^{+\infty}  f(y) \int_{-\infty}^{+\infty} 2 x y g( x - y ) \diff x \diff y \nonumber \\ & - \int_{-\infty}^{+\infty}  f(y) \int_{-\infty}^{+\infty} y^2 g( x - y ) \diff x \diff y.
\end{align}
\begin{align}
\int_{-\infty}^{+\infty}  x^2 h(x) \diff x & = \int_{-\infty}^{+\infty}  f(y) \int_{-\infty}^{+\infty}  ( x^2 - 2xy +y^2 ) g(x-y) \diff x \diff y \nonumber \\
					               & + \int_{-\infty}^{+\infty}  f(y) \int_{-\infty}^{+\infty} 2 x y g( x - y ) \diff x \diff y \nonumber \\
					               & - \int_{-\infty}^{+\infty}  f(y) \int_{-\infty}^{+\infty} y^2 g( x - y ) \diff x \diff y \label{eq:terms}
\end{align}
Eq.~\eqref{eq:terms} has three terms that we calculate separately.
\begin{align}
A_1 & = \int_{-\infty}^{+\infty}  f(y) \int_{-\infty}^{+\infty}  ( x^2 - 2xy +y^2 ) g(x-y) \diff x \diff y \nonumber \\
       & =  \int_{-\infty}^{+\infty}  f(y) \int_{-\infty}^{+\infty}  ( x-y )^2 g(x-y) \diff x \diff y \nonumber \\ 
       & = \int_{-\infty}^{+\infty}  f(y) \int_{-\infty}^{+\infty}  u^2 g(u) \diff u \diff y \nonumber \\ 
\end{align}
Now, the integral in $\diff u$ equals to:
\begin{align}
 \int_{-\infty}^{+\infty}  u^2 g(u) \diff u = \textrm{Var}[g(x)] + \textrm{E}[g(x)]^2,
\end{align}
and likewise $A_1$ becomes:
\begin{align}
A_1 & = \int_{-\infty}^{+\infty}  f(y) ( \textrm{Var}[g(x)] + \textrm{E}[g(x)]^2 ) \diff y \nonumber \\ 
	& = \textrm{Var}[g(x)] + \textrm{E}[g(x)]^2. 
\end{align}
\begin{align}
A_2  = & \int_{-\infty}^{+\infty}  f(y) \int_{-\infty}^{+\infty} 2 x y g( x - y ) \diff x \diff y \nonumber \\
	 = & 2 \int_{-\infty}^{+\infty}  y f(y) \int_{-\infty}^{+\infty} x g( x - y ) \diff x \diff y \nonumber \\ 
	 = & 2 \int_{-\infty}^{+\infty}  y f(y) \int_{-\infty}^{+\infty} (u+y) g( u ) \diff u \diff y \nonumber \\ 
	 = & 2 \int_{-\infty}^{+\infty}  y f(y) \int_{-\infty}^{+\infty} u g( u ) \diff u \diff y \nonumber \\ &+ 2 \int_{-\infty}^{+\infty}  y^2 f(y) \int_{-\infty}^{+\infty} g( u ) \diff u \diff y \nonumber \\ 
	 = & 2 \int_{-\infty}^{+\infty}  y f(y) \textrm{E}[g(x)] \diff y +  2 \int_{-\infty}^{+\infty}  y^2 f(y) \diff y \nonumber \\ 
	 = & 2 \textrm{E}[f(x)]  \textrm{E}[g(x)] + 2 ( \textrm{Var}[f(x)]   + \textrm{E}[f(x)]^2 )
\end{align}

\begin{align}
A_3 & = - \int_{-\infty}^{+\infty}  y^2 f(y) \int_{-\infty}^{+\infty}  g(x-y) \diff x \diff y  \nonumber \\ 
	& = - \int_{-\infty}^{+\infty}  y^2 f(y) \int_{-\infty}^{+\infty}  g(u) \diff u \diff y  \nonumber \\ 
	& = - \int_{-\infty}^{+\infty}  y^2 f(y) \diff y  \nonumber \\ 
	& = -( \textrm{Var}[f(x)] + \textrm{E}[f(x)]^2 )
\end{align}

Plugging the results for $A_1$, $A_2$ and $A_3$ in eq.~\eqref{eq:terms} one has:
\begin{align}
\int_{-\infty}^{+\infty}  x^2 h(x) \diff x  = & A_1 + A_2 + A_3 \nonumber \\
					        = & \textrm{Var}[g(x)] + \textrm{E}[g(x)]^2 \nonumber \\
					       & + 2 \textrm{E}[f(x)]  \textrm{E}[g(x)] \nonumber \\ & + 2 ( \textrm{Var}[f(x)] + \textrm{E}[f(x)]^2 ) \nonumber \\
					       & -( \textrm{Var}[f(x)] + \textrm{E}[f(x)]^2 )	\nonumber \\
					        = & \textrm{Var}[f(x)] 	+ \textrm{Var}[g(x)] \nonumber \\ & + ( \textrm{E}[f(x)] + \textrm{E}[g(x)] )^2.  	       
\end{align}
Finally the variance of $h(x)$ becomes:
\begin{align}
\textrm{Var}[h(x)]  & = \int_{-\infty}^{+\infty}  x^2 h(x) \diff x -  \textrm{E}[h(x)]^2 \nonumber \\
				& =  \textrm{Var}[f(x)] 	+ \textrm{Var}[g(x)]. \label{eq:res2}
\end{align}
Of course, the result of eq.~\eqref{eq:res2} can be generalized to the case of a convolution of three, four, etc. distributions. 
For the case of $(S_n*B)(x)$ we have:
\begin{align}
\text{Var}[( S_n * B )(x)]  \ = \ & \sigma_0^2 + n  \sigma_G^2, 
\end{align}
since there are $n$ convolutions of $S(x)$ and a last one with the pedestal $B(x)$.



\begin{thebibliography}{00}


\bibitem{leo} W. R. Leo, \emph{Techniques for Nuclear and Particle Physics Experiments: A How-to Approach}, 2nd ed. Springer, Berlin, Germany (1994).

\bibitem{infn}  \emph{Neutrinos: JUNO experiment debuts with extremely high precision}. Accessed on the 22/09/2026 at:  
\href{https://www.infn.it/en/neutrinos-juno-experiment-debuts-with-extremely-high-precision/}{https://www.infn.it/en/neutrinos-juno-experiment-debuts-with-extremely-high-precision}.

\bibitem{juno} \textsc{JUNO} collaboration, \emph{Potential to identify neutrino mass ordering with reactor antineutrinos at JUNO},
\href{https://iopscience.iop.org/article/10.1088/1674-1137/ad7f3e}{\emph{Chin. Phys. C} {\bf 49} (2025) 033104}
[\href{https://arxiv.org/abs/2405.18008}{\texttt{arXiv:2405.18008}}].

\bibitem{juno1} \textsc{JUNO} collaboration, \emph{Sub-percent precision measurement of neutrino oscillation parameters with JUNO}, 
\href{https://iopscience.iop.org/article/10.1088/1674-1137/ac8bc9}{\emph{Chin. Phys. C} {\bf 46} (2022) 123001}
[\href{https://arxiv.org/abs/2204.13249}{\texttt{arXiv:2204.13249}}].

\bibitem{juno2} \textsc{JUNO} collaboration, \emph{JUNO sensitivity to $^7$Be, $pep$, and CNO solar neutrinos},
\href{https://iopscience.iop.org/article/10.1088/1475-7516/2023/10/022}{\emph{JCAP} {\bf 10} (2023) 022} 
[\href{https://arxiv.org/abs/2303.03910}{\texttt{arXiv:2303.03910}}].

\bibitem{juno5} \textsc{JUNO} collaboration, \emph{JUNO sensitivity to low energy atmospheric neutrino spectra}, 
\href{https://link.springer.com/article/10.1140/epjc/s10052-021-09565-z}{\emph{Eur. Phys. J. C}  {\bf 81} (2021) 887}
[\href{https://arxiv.org/abs/2103.09908}{\texttt{arXiv:2103.09908}}].

\bibitem{juno6} \textsc{JUNO} collaboration, \emph{Prospects for geoneutrino detection with JUNO}, \href{https://arxiv.org/abs/2511.07227}{\texttt{arXiv:2511.07227}}.

\bibitem{juno7} \textsc{JUNO} collaboration, \emph{Real-time monitoring for the next core-collapse supernova in JUNO}, 
\href{https://iopscience.iop.org/article/10.1088/1475-7516/2024/01/057/pdf}{\emph{JCAP} {\bf 01} (2024) 057}
[\href{https://arxiv.org/abs/2309.07109}{\texttt{arXiv:2309.07109}}].
 
 \bibitem{res} \textsc{JUNO} collaboration, \emph{Measurement of reactor neutrino oscillation with the first JUNO data}, 
\href{https://www.nature.com/articles/s41586-026-10538-z}{\emph{Nature} {\bf 654} (2026) 343}
 [\href{https://arxiv.org/abs/2511.14593}{\texttt{arXiv:2511.14593}}].
 
\bibitem{hama}  \emph{Photomultiplier tube R5912}. Accessed on the 22/09/2026 at:  
\href{https://hep.hamamatsu.com/eu/en/products/R5912.html}{https://hep.hamamatsu.com/eu/en/products/R5912.html}.

\bibitem{hama2} Hamamatsu Photonics K. K., \emph{Photomultiplier tubes: Basics and Applications}, 4th edition, Hamamatsu Photonics K. K. (2017). Accessed on the 22/09/2026 at:
\href{https://www.hamamatsu.com/content/dam/hamamatsu-photonics/sites/documents/99_SALES_LIBRARY/etd/PMT_handbook_v4E.pdf}
{https://www.hamamatsu.com/content/dam/hamamatsu-photonics/sites/documents/99\_SALES\_LIBRARY/etd\\/PMT\_handbook\_v4E.pdf}.

\bibitem{qe} Hamamatsu Photonics, \emph{Large photocathode area photomultiplier tubes}. Accessed on the 22/09/2026 at:  
\href{https://hep.hamamatsu.com/content/dam/hamamatsu-photonics/sites/documents/99_SALES_LIBRARY/etd/LARGE_AREA_PMT_TPMH1376E.pdf}
{https://hep.hamamatsu.com/content/dam/hamamatsu-photonics/sites/documents/99\_SALES\_LIBRARY/etd/LARGE\\ \_AREA\_PMT\_TPMH1376E.pdf}.

\bibitem{me1} L. N. Kalousis, \emph{Calibration of the Double Chooz detector and cosmic background studies}, \href{http://inspirehep.net/record/1295030}{\emph{PhD thesis, University of Strasbourg} (2012)}. 

\bibitem{smi1} O. Ju. Smirnov, \emph{Energy and spatial resolution of a large-volume liquid-scintillator detector}, 
\href{https://link.springer.com/article/10.1023/A:1024458203966}{\emph{Instrum. Exp. Tech.} {\bf 46} (2003) 327}
[\href{https://arxiv.org/abs/1811.02321}{\texttt{arXiv:1811.02321}}].
 
\bibitem{geant} S. Agostinelli {et al.}, \emph{Geant4 - A Simulation Toolkit},  \href{https://www.sciencedirect.com/science/article/abs/pii/S0168900203013688}{\emph{Nucl. Instrum. Meth.} {\bf A} 506 (2003) 250}.
 
\bibitem{glen} GLG4sim page. Accessed on the 22/09/2026 at:  \href{https://www.phys.ksu.edu/personal/gahs/GLG4sim/}{https://www.phys.ksu.edu/personal/gahs/GLG4sim/}.

\bibitem{pap} A. Papoulis, \emph{Probability, Random Variables and Stochastic Processes}, 4th ed. McGraw-Hill Europe (2002). 

\bibitem{Bellamy} E. H. Bellamy {et al.}, \emph{Absolute Calibration and Monitoring of a spectrometric channel using a photomultiplier}, 
\href{https://www.sciencedirect.com/science/article/pii/016890029490183X}{\emph{Nucl. Instrum. Meth. A} {\bf 339} (1994) 468}. 

\bibitem{pois2} Y. Hu {et al.}, \emph{On the Poisson approximation to photon distribution for faint lasers}, 
\href{https://www.sciencedirect.com/science/article/abs/pii/S0375960107003660}{\emph{Physics Letters} {\bf A 367} (2007) 173}
[\href{https://arxiv.org/abs/math-ph/0609063}{\texttt{arXiv:0609063}}].

\bibitem{rade} J. Rademacker, \emph{An exact formula to describe the amplification process in a photomultiplier tube}, 
\href{https://www.sciencedirect.com/science/article/abs/pii/S0168900201020551}{ \emph{Nucl. Instrum. Methods A} {\bf 484} (2002) 432}
[\href{https://arxiv.org/abs/physics/0406036}{\texttt{arXiv:0406036}}].

\bibitem{DCID1} E. Calvo {et al.}, \emph{Characterization of large area photomutipliers under low magnetic fields: design and performances of the magnetic shielding for the Double Chooz neutrino experiment}, 
\href{https://www.sciencedirect.com/science/article/pii/S016890021001199X}{\emph{Nucl. Instrum. Meth.} {\bf A 621} (2010) 222} 
 [\href{https://arxiv.org/abs/0905.3246}{\texttt{arXiv:0905.3246}}]. 

\bibitem{DCID2} C. Bauer {et al.}, \emph{Qualification Tests of 474 Photomultiplier Tubes for the Inner Detector of the Double Chooz Experiment}, 
\href{https://iopscience.iop.org/article/10.1088/1748-0221/6/06/P06008}{\emph{JINST} {\bf 6} (2011) P06008}  [\href{https://arxiv.org/abs/1104.0758}{\texttt{arXiv:1104.0758}}]. 

\bibitem{DCID3} T. Matsubara {et al.}, \emph{Evaluation of 400 low background 10-in. photo-multiplier tubes for the Double Chooz experiment}, 
\href{https://www.sciencedirect.com/science/article/pii/S016890021101775X?via\%3Dihub}{\emph{Nucl. Instrum. Meth.} {\bf A 661} (2012) 16}. 
[\href{https://arxiv.org/abs/1104.0786}{\texttt{arXiv:1104.0786}}]. 

\bibitem{me2} L. N. Kalousis {et al.}, \emph{A fast numerical method for photomultiplier calibration},
 \href{https://iopscience.iop.org/article/10.1088/1748-0221/15/03/P03023}{\emph{JINST} {\bf 15} (2020) P03023}
 [\href{https://arxiv.org/abs/1911.06220}{\texttt{arXiv:1911.06220}}]. 
 
 \bibitem{Smirnov}  O. Ju. Smirnov {et al.}, \emph{Methods for precise photoelectron counting with photomultipliers}, 
\href{https://www.sciencedirect.com/science/article/pii/S0168900200003375}{\emph{Nucl. Instrum. Meth.} {\bf A 451} (2000) 623}. 

\bibitem{error} M. Abramowitz and I. A. Stegun, \emph{Handbook of Mathematical Functions: with Formulas, Graphs, 
and Mathematical Tables}, Dover Publications, 0009-Revised edition (June 1, 1965).  

 \bibitem{ana}  L. N. Kalousis, \emph{An analytical model for photomultiplier tube calibration}.
 \href{https://www.sciencedirect.com/science/article/abs/pii/S0168900223009439}{\emph{Nucl. Instrum. Meth.} {\bf A 1058} (2024) 168943} 
[\href{https://arxiv.org/abs/2304.08735}{\texttt{arXiv:2304.08735}}]. 

\bibitem{salda} R. Saldanha et al., \emph{Model independent approach to the single photoelectron calibration of photomultiplier tubes}, 
 \href{https://www.sciencedirect.com/science/article/abs/pii/S016890021730311X}{\emph{Nucl. Instrum. Meth.} {\bf A 863} (2017) 35} 
 [\href{https://arxiv.org/abs/1602.03150}{\texttt{arXiv:1602.03150}}]. 
 
\bibitem{dark} {DarkSide} collaboration, \emph{Light Yield in DarkSide-10: a Prototype Two-phase Argon TPC for Dark Matter Searches}, 
\href{https://www.sciencedirect.com/science/article/pii/S0927650513001254?via\%3Dihub}{\emph{Astropart. Phys.} {\bf 49} (2013) 44-51}  
[\href{https://arxiv.org/abs/1204.6218}{\texttt{arXiv:1204.6218}}]. 

\bibitem{me3} L. N. Kalousis, \emph{Calibration of photomultiplier tubes},
 \href{https://iopscience.iop.org/article/10.1088/1748-0221/18/07/P07016}{\emph{JINST} {\bf 18} (2023) P07016}
 [\href{https://arxiv.org/abs/2304.07292}{\texttt{arXiv:2304.07292}}].  
   
\bibitem{Arfken} G. B. Arfken and H. J. Weber, \emph{Mathematical Methods for Physicists}, 6th Ed. Elsevier Academic Press, San Diego, USA (2005).

\bibitem{fftw} FFTW page. Accessed on the 22/09/2026 at: \href{http://www.fftw.org}{http://www.fftw.org} 

\bibitem{root} R. Brun and F. Rademakers, \emph{ROOT - An Object Oriented Data Analysis Framework}, \emph{Proceedings AIHENP'96 Workshop}, Lausanne Switzerland (1996),
\href{https://www.sciencedirect.com/science/article/pii/S016890029700048X?via\%3Dihub}{\emph{Nucl. Instrum. Meth.} {\bf A 389 } (1997) 81.} \\
See also \href{http://root.cern.ch/}{http://root.cern.ch/}

\bibitem{git} lkalousis github. Accessed on the 22/09/2026 at: \href{https://github.com/lkalousis/PMTCalib}{https://github.com/lkalousis/PMTCalib}
 
\bibitem{perf} \textsc{JUNO} collaboration, \emph{Initial performance results of the JUNO detector}, 
\href{hhttps://iopscience.iop.org/article/10.1088/1674-1137/ae3dc1}{\emph{Chinese Phys.} {\bf C 50} (2026) 043001}
[\href{https://arxiv.org/abs/2511.14590}{\texttt{arXiv:2511.14590}}].
 
 \bibitem{wire} V. Fran\c{c}ais, \emph{Description and simulation of the physics of Resistive Plate Chambers},  
 \href{https://tel.archives-ouvertes.fr/tel-01727712}{\emph{PhD thesis, University of Clermont Auvergne} (2017)}. 
 
\bibitem{IMBcalib} \textsc{IMB} collaboration, \emph{Calibration of the IMB detector}, 
\href{https://www.sciencedirect.com/science/article/pii/0168900295900187?via\%3Dihub}{\emph{Nucl. Instrum. Meth.} {\bf A~352} (1995) 629}.

\bibitem{minuit2} F. James and M. Winkler, \emph{C++ MINUIT User's Guide},
\href{https://root.cern.ch/root/htmldoc/guides/minuit2/Minuit2.html}{https://root.cern.ch/root/htmldoc/guides/minuit2/Minuit2.html}
 
 \bibitem{IMB} \textsc{IMB} collaboration, \emph{IMB-3: a large water Cherenkov detector for nucleon decay and neutrino interactions}, 
\href{https://www.sciencedirect.com/science/article/pii/016890029390998W?via\%3Dihub}{\emph{Nucl. Instrum. Meth.} {\bf A 324} (1993) 363}.
   
\bibitem{DC} {Double Chooz} collaboration, \emph{The Double Chooz antineutrino detectors},  
 \href{https://link.springer.com/article/10.1140/epjc/s10052-022-10726-x}{\emph{Eur. Phys. J.} {\bf C} (2022) 82}
 [\href{https://arxiv.org/abs/2201.13285}{\texttt{arXiv:2201.13285}}].  
 
 \bibitem{agi}  \emph{High-performance pulse generation with precise timing, Model: 81150A}. Accessed on the 22/09/2026 at:  
 \href{https://www.keysight.com/en/pd-1287544-pn-81150A/pulse-function-arbitrary-noise-generator?nid=-536902255.748669.00&cc=FR&lc=fre}{https://www.keysight.com/en/pd-1287544-pn-81150A}.
 
 \bibitem{fib} Thorlabs fiber BFH48-600. Accessed on the 22/09/2026 at:  
 \href{https://www.thorlabs.com/thorproduct.cfm?partnumber=BFH48-600}{https://www.thorlabs.com/thorproduct.cfm?partnumber=BFH48-600}
 
 \bibitem{lecroy} Teledyne Lecroy,  WavePro 7 Zi-A Series. Accessed on the 22/09/2026 at:  
 \href{http://cdn.teledynelecroy.com/files/pdf/wavepro_7_zi-a_datasheet.pdf}{http://cdn.teledynelecroy.com/files/pdf/wavepro\_7\_zi-a\_datasheet.pdf}
    
\bibitem{icecube} {IceCube} collaboration, \emph{In-situ calibration of the single-photoelectron charge response of the IceCube photomultiplier tubes}, 
 \href{https://iopscience.iop.org/article/10.1088/1748-0221/15/06/P06032}{\emph{JINST} {\bf 15} (2020) P06032}   
 [\href{https://arxiv.org/abs/2002.00997}{\texttt{arXiv:2002.00997}}].  

\end{thebibliography}


\section*{Acknowledgements}
The data used in this publication were taken in the laboratory of Dr. M. A. Dracos and we wish to thank him for allowing us to use them for the purposes of this communication.

\end{document}